\documentclass[a4paper,11pt]{article}
\pdfoutput=1 

\usepackage{jheppub} 

\usepackage[T1]{fontenc} 
\usepackage{orcidlink,booktabs}
\usepackage{soul,amsmath,amssymb}
\usepackage{siunitx,bm}
\usepackage{comment}

\usepackage{lettrine}
\usepackage{Zallman}

\DeclareFontFamily{U}{wncy}{}
    \DeclareFontShape{U}{wncy}{m}{n}{<->wncyr10}{}
    \DeclareSymbolFont{mcy}{U}{wncy}{m}{n}
    \DeclareMathSymbol{\Sh}{\mathord}{mcy}{"58}

\title{\boldmath Super-Nyquist ultralight dark matter searches with broadband atom gradiometers}

\author[]{Leonardo Badurina\,\orcidlink{0000-0003-4575-5127},}
\author[]{Ankit Beniwal\,\orcidlink{0000-0003-4849-0611} and}
\author[]{Christopher McCabe\,\orcidlink{0000-0002-4767-821X}}

\affiliation[]{Theoretical Particle Physics and Cosmology (TPPC), Department of Physics, \\King’s College London, Strand, London, WC2R 2LS, UK}

\emailAdd{leonardo.badurina@kcl.ac.uk}

\abstract{
Atom gradiometers have emerged as compelling broadband probes of scalar ultralight dark matter (ULDM) candidates that oscillate with frequencies between approximately $10^{-2}$~Hz and $10^3$~Hz.  ULDM signals with frequencies greater than $\sim 1$~Hz exceed the expected Nyquist frequency of atom gradiometers, and so are affected by aliasing and related phenomena, including signal folding and spectral distortion. To facilitate the discovery of super-Nyquist ULDM signals, in this work we investigate the impact of these effects on parameter reconstruction using a robust likelihood-based framework. We demonstrate that accurate reconstruction of ULDM parameters can be achieved as long as the experimental frequency resolution is larger than the ULDM signal linewidth. Notably, as ULDM candidates whose frequencies differ by integer multiples of the sampling frequency are identified at the same aliased frequency, our discovery analysis recovers discrete islands in parameter space. Our study represents the first comprehensive exploration of aliasing in the context of dark matter direct detection and paves the way for enhanced ULDM detection strategies with atom gradiometers.}


\arxivnumber{}

\preprint{KCL-PH-TH/2023-30, AION-REPORT/2023-05}

\begin{document} 

\maketitle
\flushbottom

\section{Introduction}\label{sec: intro}

\lettrine{E}ver since the formulation of the dark matter (DM) hypothesis, the search for dark matter in direct detection experiments has been one of the greatest priorities in particle physics~\cite{APPEC, Cooley:2022ufh}. Until recently, the possibility of charting the strikingly diverse and vast landscape of DM models beyond conventional GeV-scale candidates seemed like a remote possibility. Now, thanks to extraordinary advancements in a wealth of cutting-edge technologies with ever-increasing sensitivity to minute effects, it is expected that large regions of DM model space will be within the reach of the next generation of direct detection experiments, such as atom interferometers.

In addition to being excellent probes of gravitational waves in the mid-frequency gap~\cite{Badurina:2021rgt},
large-scale atom interferometer experiments, such as AION~\cite{Badurina:2019hst}, MAGIS~\cite{MAGIS-100:2021etm}, MIGA~\cite{Canuel:2017rrp}, ELGAR~\cite{Canuel:2019abg}, and ZAIGA~\cite{Zhan:2019quq}, would be powerful probes of ultralight dark matter (ULDM). In particular, these experiments would be especially ideal probes of scalar ULDM signatures through their exquisite sensitivity to changes in atomic structures. Indeed, scalar ULDM with dilatonic couplings to Standard Model (SM) operators would give rise to time-varying oscillations in atomic transition frequencies~\cite{Stadnik:2015kia, Arvanitaki:2014faa}, which in turn would generate an oscillatory non-vanishing phase difference between pairs of spatially-separated interferometers that are interrogated by the same set of lasers through a gradiometer configuration~\cite{Arvanitaki:2016fyj, Badurina:2021lwr}. 

As first shown in Ref.~\cite{Arvanitaki:2016fyj}, and later studied in detail within the context of the AION and MAGIS experiments, terrestrial long-baseline single-photon vertical atom gradiometers and space-based experiments that operate in broadband mode would be especially powerful probes of scalar ULDM with masses between $\sim 10^{-18}$~eV and $10^{-13}$~eV, corresponding to signals oscillating at $\sim10^{-2}$~Hz and $\sim10^3$~Hz, respectively. Importantly, these experiments are expected to outcompete other complementary probes in this frequency range, such as atomic clocks~\cite{Arvanitaki:2014faa}, the MICROSCOPE  experiment~\cite{Berge:2017ovy}, torsion balance experiments~\cite{Wagner:2012ui}, the AURIGA experiment~\cite{Branca:2016rez} and superradiance constraints from the observations of fast-spinning stellar-mass black-holes in X-ray binaries~\cite{Baryakhtar:2020gao}. 

The high-frequency range of broadband interferometer experiments, by which we mean the window containing signal frequencies greater than $\sim 1$~Hz, offers particularly interesting search prospects in light of projected exclusion limits from future experiments, and theoretical considerations. For instance, future non-interferometer experiments, such as resonant cavity experiments~\cite{Geraci:2018fax}, will only set weak limits in the 1--100 Hz frequency window compared with those arising from proposed vertical long-baseline gradiometers operating single-photon atomic transitions, e.g., AION and MAGIS. Hence, it is expected that only long-baseline interferometers will be able to set constraints on non-conventional scalar dark matter production mechanisms in this mass window, such as the thermal misalignment mechanism~\cite{Batell:2021ofv}, and probe characteristic signatures of ULDM in direct detection experiments that are most amplified in this frequency range, such as gravitational focusing~\cite{Kim:2021yyo}. Moreover, the high-frequency region of parameter space accessible to long-baseline gradiometers is not fine-tuned, which is often considered as an important criterion for selecting a region of parameter space to target. Assuming a UV-cutoff at 10~TeV, loop-corrections to the mass of a scalar ULDM candidate with linear couplings to electron masses would be smaller than the renormalised mass for couplings satisfying the relation $d_\phi \lesssim m_\phi/(3.3\times 10^{-10}~\mathrm{eV})$, which would be within the reach of a 1-km gradiometer operating with the parameters proposed in Ref.~\cite{Badurina:2022ngn}.\footnote{The UV-cutoff depends on the scale of new physics, which could lie below 10~TeV. For instance, assuming Higgs-portal DM, the cut-off would be set by the Higgs mass; assuming couplings at tree-level to only the electron mass, the cut-off would instead be set by the electron mass.~While both models are feasible, we stress that the latter would require elaborate and \emph{ad hoc} model-building.} 

Aside from these considerations, the projected reach of long-baseline atom interferometers may start to become limited by mass-density fluctuations induced by seismic activity below $\sim0.5$~Hz; in turn, this means that the peak sensitivity of a 1-km gradiometer operating with the parameters proposed in Ref.~\cite{Badurina:2022ngn} would be shifted from 0.1~Hz to $\sim 1$~Hz, which lies close to or beyond the expected sampling frequency of these experiments. As is well known in signal analysis, any signal oscillating at a frequency larger than half of the experimental sampling frequency, also known as the Nyquist frequency, is aliased to a lower frequency between zero and the Nyquist frequency, see e.g., Refs.~\cite{10.5555/227373, EPFL}.~Additionally, important spectral distortions due to aliasing would affect the qualitative features of a putative signal, and thus impact the ability to correctly identify a signal as being of a ULDM origin, as briefly discussed first in Ref.~\cite{Derevianko:2016vpm} within the context of ULDM searches with a network of atomic clocks. As a study in this direction is lacking within the context of ULDM searches with atom interferometers, here we provide a complete and versatile likelihood-based analysis framework, which makes use of the machinery developed in Refs.~\cite{Badurina:2022ngn, Foster:2017hbq}, for discovering these high-frequency signals. With the aid of Monte Carlo (MC) simulations, we also confirm the robustness of our analysis strategy in reconstructing the properties of an injected signal. Importantly, leveraging on these statistical tools, we provide pre- and post-data collection strategies that broadband atom interferometer experiments should consider to maximise the potential to discover a super-Nyquist ULDM signal.

The rest of the paper is organised as follows. In section~\ref{sec: ULDM signal}, we review the scalar ULDM signal in broadband vertical atom gradiometers employing single-photon atomic transitions. In section~\ref{sec: freq-dom analysis}, we present the likelihood-based analysis of a ULDM signal in the frequency domain. In section~\ref{sec: high-freq analysis}, we provide a detailed discussion of aliasing within the context of ULDM searches: in section~\ref{subsec: intro aliasing}, we provide the reader with an overview of the concepts and phenomena associated with aliasing, such as Nyquist windows and folding; in sections~\ref{subsec: alias vs nonalias}-\ref{subsec: spectral distortions} we present several techniques to correctly reconstruct super-Nyquist ULDM signals that are affected by aliasing; and in section~\ref{subsec: example search} we apply these techniques to a discovery search based on several MC realisations of the data.~Finally, we summarise our results in section~\ref{sec: summary}.~Several appendices provide further details that complement and validate our analysis and results. 

\section{Scalar ULDM signal in vertical atom gradiometers}\label{sec: ULDM signal}

In light of its large occupation number, small mean velocity and velocity dispersion that is characteristic of DM in the Milky Way, scalar ULDM can be modeled as a temporally and spatially oscillating, non-relativistic classical field with frequency largely set by the DM mass, $m_\phi = 2\pi f_\phi$, and small kinetic corrections~\cite{ Ferreira:2020fam}.\footnote{In this work, we will mainly quote masses in Hz and eV. The conversion between the two quantities is given by $f_\phi = m_\phi / (4.136 \times 10^{-15}~\mathrm{eV})$~Hz.} As a result of the wave's dispersion $\Delta \omega_\phi = 2\pi \Delta f_\phi \sim m_\phi v_0 \sigma_v$ where $\{v_0,~\sigma_v \} \sim 10^{-3}$, ULDM is characterised by a coherence time $\tau_c \equiv 2 \pi / (m_\phi \, v_0^2)$, which sets the timescale on which the field amplitude and phase vary considerably. Assuming the random phase model, the ULDM field can be expressed as a sum of its Fourier components with uncorrelated phases~\cite{Hui:2021tkt}.

In the context of direct detection experiments, which aim to measure a time-varying signal that is proportional to the ULDM field itself over an integration time $T_\mathrm{int}$, it is advantageous to express the ULDM field in terms of an experiment's frequency resolution $\Delta f \equiv 1/T_\mathrm{int}$. Without loss of generality, assuming the random phase model and neglecting its spatial variation, a scalar ULDM field can be expressed as \cite{Badurina:2022ngn}
\begin{equation}
\begin{gathered}
    \phi(t) = \frac{\sqrt{\rho_\mathrm{DM}}}{m_\phi} \sum_{a}{\alpha_a \sqrt{F_\mathrm{DM}(v_a)}\cos{\left (\omega_a t + \theta_a \right)}} \, , \\
    F_\mathrm{DM}(v_a) = \int_{v_a \, - \, \Delta v/2}^{v_a \, + \, \Delta v/2} dv \, f_\mathrm{DM}(v) \, ,
\end{gathered} 
\label{eq:full-field}
\end{equation}
where $\rho_\mathrm{DM} = 0.3$\,GeV/cm$^3$ is the local DM density, $\omega_a \simeq m_\phi \, (1+v_a^2/2)$ is the angular frequency of the ULDM wave for a given speed $v_a$ and $\theta_a \in [0,2\pi)$ is a random phase.~The sum is over velocity classes that are identified by the index $a$; the size of each velocity class $\Delta v$ is related to the experiment's frequency resolution $\Delta f = 1/T_\mathrm{int} \simeq m_\phi v_0 \Delta v/2\pi$.\footnote{The former equality follows from the discrete Fourier transform of the data, whereas the latter follows from the signal's kinetic energy.}~The variable $\alpha_a$ is Rayleigh distributed with $\langle \alpha_a^2 \rangle  = 2$; its probability density function is given by
\begin{equation}
    P(\alpha_a) = \alpha_a \exp \left ( - \frac{\alpha_a^2}{2} \right ) \, .
    \label{eq: Rayleigh distribution}
\end{equation}
The DM speed distribution is denoted by $f_{\rm{DM}} (v)$, which we assume to correspond to the Standard Halo Model (SHM)~\cite{Lewin:1995rx, Drukier:1986tm}, namely
\begin{align}\label{eqn:SHM_dist}
    f_{\rm{DM}} (v) = \frac{v}{ \sqrt{2 \pi} \sigma_v v_{\mathrm{obs}}} & e^{-\left(v+v_{\mathrm{obs}}\right)^{2}/ (2 \sigma_v^2)} \left(e^{4 v v_{\mathrm{obs}} / (2 \sigma_v^2)}-1\right)\, ,
\end{align}
where $\sigma_v$ is the velocity dispersion which is set, at the solar position, by the value of the local standard of rest $v_0= \sqrt{2 }\sigma_v \approx \SI{238}{km/s}$, and $v_\mathrm{obs} \approx \SI{252}{km/s}$ is the average speed of the Earth relative to the halo rest frame~\cite{Baxter:2021pqo}.\footnote{Although the DM speed distribution is characterised by a cut-off at the escape velocity $v_\mathrm{esc} \sim \SI{800}{km/s}$ in the Earth's frame and may feature substantial radial anisotropic components, commonly referred to as the Gaia Sausage or Gaia-Enceladus~\cite{Evans:2018bqy}, we expect the simple SHM form in Eq.~\eqref{eqn:SHM_dist} with no cut-off and anisotropies to be sufficient for the analysis strategy presented here.}

We end this discussion by noting that the spatial structure of the ULDM wave can be neglected in individual gradiometer experiments by observing that: \textit{(i)} on average the magnitude of wave vector $k_\phi = m_\phi v = 1/\lambda_\mathrm{dB}$, where $\lambda_\mathrm{dB}$ is the DM's de Broglie wavelength, is suppressed by a factor of $v \sim 10^{-3}$ relative to the angular frequency; $\textit{(ii)}$ the longest length scale in an atom gradiometer experiment is set by the length of the baseline, which, in the mass range of interest, is significantly smaller than~$\lambda_\mathrm{dB}\simeq\left( 10^{-14}\,\mathrm{eV}/m_{\phi}\right) 2\times 10^7 \,\mathrm{km}$. Hence, corrections to the phase from the ULDM spatial structure are highly subdominant.\footnote{The spatial structure of the ULDM wave could be relevant when performing ULDM searches with \textit{networked} atom gradiometers. When the distance between gradiometer experiments (e.g.\ MAGIS and AION) is comparable to the field's de Broglie wavelength, the correction to the ULDM's phase would be on the order of the time-dependent phase~\cite{Foster:2020fln}. Hence, two networked experiments that are located at antipodal points could be used to explore the spatial structure of the ULDM scalar candidates with masses $m_\phi \gtrsim 10^{-11}$~eV. A study on the physics potential of such searches is beyond the scope of this work.}

\subsection{ULDM-induced differential phase shift}\label{subsec: ULDM phase shift}

\begin{figure}[t]
    \centering
    \includegraphics[width = \textwidth]{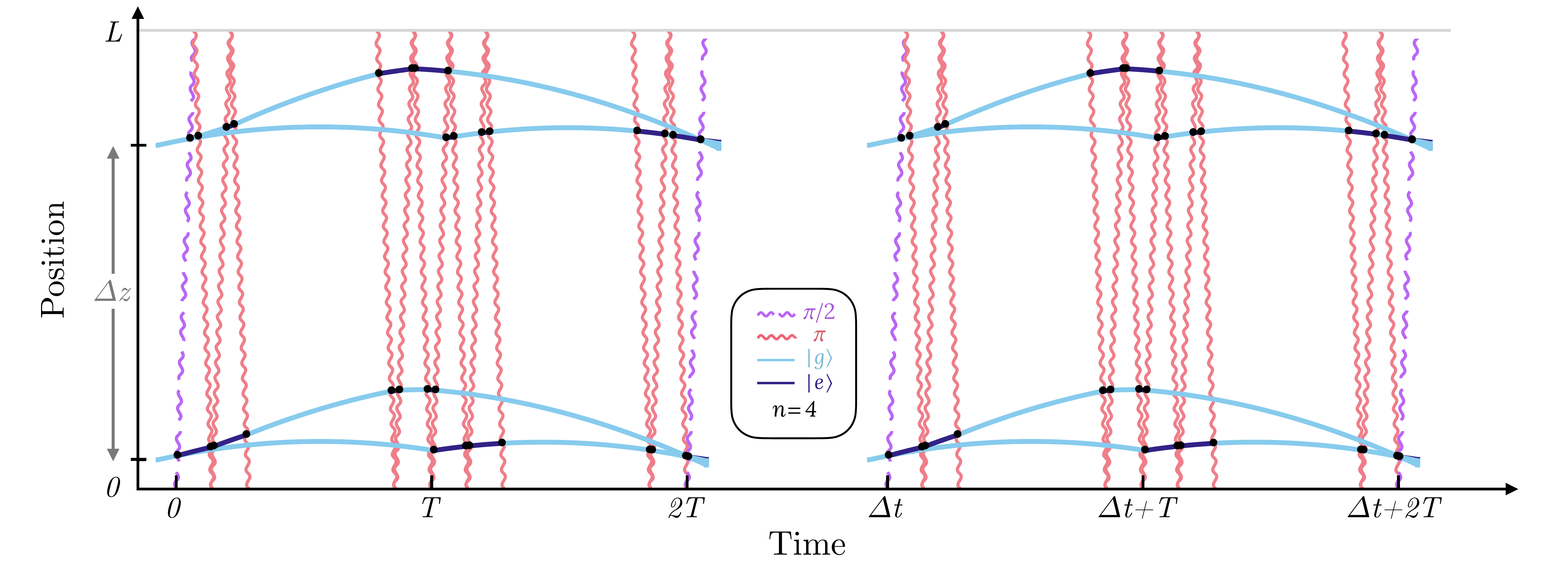} \\
    \caption{
    Schematic representation of the atom gradiometer sequence considered in this work for $n=4$ LMT kicks.
    The atom's excited ($| e \rangle$) and ground ($| g \rangle$) states are shown in blue and cyan, respectively. 
$\pi/2$- and $\pi$-pulses are displayed as wavy lines in fuchsia (dashed) and red (solid), respectively. 
Atom-light interactions are indicated with black dots.
The length of the baseline is $L$, the distance between the atom interferometers is $\Delta z$, and $T$ is the interrogation time. 
Here, we show two sequences, which are temporally separated by the sampling interval $\Delta t$.
 }
    \label{fig: Scheme example}
\end{figure}

Linear interactions between scalar ULDM and SM photons/electrons give rise to the oscillation of the electron mass $m_e$ and fine-structure constant $\alpha$~\cite{Damour:2010rp, Stadnik:2014tta}, 
\begin{align}
	m_e (t) &= m_e \left[1 + d_{m_e} \sqrt{4 \pi G_N} \, \phi(t) \right]\,, \\
    \alpha (t) &\approx \alpha \left[1 + d_e \sqrt{4 \pi G_N}\,  \phi(t) \, \right] \,,	
\end{align}
where the parameters $\{d_e,\,d_{m_e}\}$ denote the coupling strengths relative to the Planck mass, which we explicitly express in terms of Newton's gravitational constant $G_N$. These couplings lead to oscillations in atomic transition frequencies which could be detected in the differential phase shift $\Phi$ measured between two atom interferometers that compose an atom gradiometer. Explicitly, as first calculated in Ref.~\cite{Arvanitaki:2016fyj} and subsequently generalised to baselines of arbitrary length in Refs.~\cite{Badurina:2021lwr, Badurina:2022ngn}, the ULDM gradiometer phase shift measured after a time $m\Delta t$ from the start of the first experiment between two coupled interferometers, which are separated by a distance $\Delta z$, takes the following form:
\begin{equation}\label{eq:ULDMphase}
   \Phi_{\mathrm{DM}, m} = \frac{\Delta z}{L} \sum_{a} \alpha_a \sqrt{F_{\rm{DM}} (v_a)} \, A_a \cos \phi_{a,m}\;,
\end{equation}
where 
\begin{gather}\label{eq:Aa def}
    A_{a} = \frac{8}{\sqrt{2}} \frac{\overline{\Delta \omega_{A}}}{\omega_{a}} \sin \left[\frac{\omega_{a} n L}{2}\right] \sin \left[\frac{\omega_{a} T}{2}\right] \sin \left[\frac{\omega_{a}\left(T-(n-1) L\right)}{2}\right] \,, \\
    \overline{\Delta \omega_A} =  d_{\phi} \sqrt{4\pi G_N} \frac{ \sqrt{\rho_{\mathrm{DM}}}}{m_{\phi}} \, \omega_A \,.
\end{gather}
Here, we have assumed the `broadband' atom gradiometer sequence shown in Fig.~\ref{fig: Scheme example}, 
which consists of an initial $\pi/2$-pulse, $(4n - 3)$ $\pi$-pulses from alternating directions, and a final $\pi$/2-pulse, where $n$ is the number of large momentum transfer (LMT) kicks.

In eq.~\eqref{eq:ULDMphase}, $\phi_{a,m} \supset \omega_a m \Delta t + \theta_a$ is the phase of the DM wave at the end of the interferometer sequence; the amplitude $A_a$ depends both on experimental parameters (the number of LMT kicks $n$, the interrogation time $T$, the length of the baseline $L$ characterising the interferometric sequence, and the angular frequency $\omega_A$ of the optical transition\footnote{For the $5\mathrm{s}^2 \,^1\mathrm{S}_0\leftrightarrow 5\mathrm{s}5\mathrm{p} \,^3\mathrm{P}_1$ clock transition 
in $^{87}\mathrm{Sr}$, which we assume throughout this work, $\omega_A=2.697\times10^{15}~\mathrm{rad}/\mathrm{s}$.}) and phenomenological parameters (the local dark matter density $\rho_\mathrm{DM}$, the ULDM mass $m_\phi$, and the ULDM-SM coupling strength $d_\phi \in \{d_e,\,d_{m_e}\}$).~In this work, we assume the parameters listed in Table~\ref{table: experimental parameters}.

The dependence of Eq.~\eqref{eq:ULDMphase} on the random and uncorrelated variables $\alpha_a$ and $\theta_a$ implies that the ULDM signal is stochastic~\cite{Centers:2019dyn}. In the next section, we review the details regarding the characterisation of stochastic signals in the frequency domain, and we show how a likelihood-based framework can be used to search for a ULDM signal. This framework was developed in Ref.~\cite{Badurina:2022ngn} and builds on previous work in the context of axion-like particle searches~\cite{Foster:2017hbq, Foster:2020fln}.

\section{Frequency-domain ULDM signal analysis}\label{sec: freq-dom analysis}

The data that is measured by an interferometer consists of a finite number of phase measurements at discrete points in time, and hence constitutes a time-series of finite length. Here, we assume a constant time interval $\Delta t$ between successive measurements. As a DM signal is characterised by a frequency largely set by its mass and a frequency spread dictated by its speed distribution, the appropriate tool for analysing the data is the discrete power spectral density (PSD), which is defined as 
\begin{equation}\label{eqn:PSD_def}
    S_{\mathrm{DM}, k} \equiv \frac{(\Delta t)^2}{T_\mathrm{int}}\left |\widetilde{\Phi}_{\mathrm{DM}, k}\right|^2 \, , 
\end{equation}
where $\widetilde{\Phi}_{\mathrm{DM}, k}$ is the discrete Fourier transform of the data, $k \in \{0,1,...,N-1\}$ labels the frequency bin accessible to the experiment, and $N = T_\mathrm{int}/\Delta t$. As derived in Ref.~\cite{Badurina:2022ngn}, without loss of generality, the expectation value of the signal's PSD takes the form
\begin{equation}\label{eqn:exp_PSD}
    \langle  S_{\mathrm{DM}, k} \rangle = \frac{\pi}{2} \left (\frac{\Delta z}{L} \right)^2 A_{k}^2 \frac{1}{\Delta \omega}\int_{\omega_k \, - \, \Delta \omega/2}^{\omega_k \, + \, \Delta \omega/2} d\omega \, \frac{f_\mathrm{DM}(v_\omega)}{m_\phi \, v_\omega} \, ,
\end{equation}
where we define $v_\omega = \sqrt{2\omega/m_\phi-2}$\,, and express the $k^\mathrm{th}$ angular frequency and resolution as $\omega_k = 2\pi k/T_\mathrm{int}$ and $\Delta \omega = 2 \pi/T_\mathrm{int}$, respectively. Here, $A_k$ is as defined in Eq.~\eqref{eq:Aa def} under the redefinition $\omega_a \rightarrow \omega_k$.

By construction, the observable spectral content of the ULDM-induced signal depends on the duration of the experiment relative to the signal's coherence time, $T_{\rm{int}}/\tau_c$. When $T_{\rm{int}}/\tau_c \ll 1$, the size of the frequency bins exceed the signal's spectral linewidth $\Delta f_\phi \approx 1/\tau_c$. In turn, this implies that the signal is observed in a single frequency bin, i.e., in Eq.~\eqref{eq:ULDMphase}, $F_\mathrm{DM} = 1$  and the sum is carried over a single frequency component, so that Eq.~\eqref{eqn:exp_PSD} is only non-zero for a single value of $k$. The ULDM field is then well-described by oscillations at $\omega_\phi \approx m_\phi$ with a fixed but random amplitude and phase, such that the signal's PSD corresponds to a single spike in the frequency bin centred at $\sim f_\phi$.~Thus, in this case, we would expect the mass resolution to be comparable to the experiment's frequency resolution, namely $\Delta f = 1/T_\mathrm{int}$. On the other hand, when $T_{\rm{int}}/\tau_c \gg 1$, the size of the frequency bins would be smaller than the signal's dispersion $(\propto 1/\tau_c$) such that the spectral content of the signal could be resolved, i.e., in this limit, $F_\mathrm{DM}(v_a) = f_\mathrm{DM}(v_a) \, \Delta v$ and the sum in Eq.~\eqref{eq:ULDMphase} is over all resolvable DM speeds~\cite{Foster:2017hbq}. In practice, to resolve the DM speed distribution which is imprinted onto the signal, it is necessary to choose a sufficiently long integration time so that the frequency resolution satisfies $\Delta f \lesssim  10^{-6} \, f_\phi$. For DM masses in the high-frequency regime considered in this paper (i.e., $f_\phi \gtrsim 1 $~Hz), this requires $T_\mathrm{int} \gtrsim 10^6$\,s. 
For book-keeping purposes, we summarise the relationship between the key parameters in the frequency and time domain in Table~\ref{tab:Dictionary}.

\begin{table}[t]
    \begin{center}
    \begin{tabular}{ 
    c c  c c c c c}
    \toprule
    $L$~[m] & $T$~[s] & $n$ & $\Delta z$~[m] & $\delta \phi \, [1/\sqrt{\mathrm{Hz}}]$ & $T_\mathrm{int}$~[s] & $f_s$~[Hz] \\
    \midrule \midrule
    $1000$ & 1.7 & $2500$ & $970$ & $ 10^{-5}$ & $10^8$ & $1$\\
    \bottomrule
    \end{tabular}
    \end{center}
    \caption{List of experimental parameters assumed in this paper. These could be implemented in future vertical gradiometers, e.g., AION-km. Here, $L$ is the length of the baseline, $T$ is the interrogation time, $4n-1$ is the total number of large momentum transfer kicks transferred during a single cycle, $\Delta z$ is the gradiometer length, $\delta \phi$ is the shot noise-limited phase resolution per interferometer, $T_\mathrm{int}$ is the total integration time for the experiment and $f_s$ is the sampling frequency.} 
    \label{table: experimental parameters}
\end{table}

\subsection{Likelihood-based analysis}\label{subsec: likelihood}

As was shown in Ref.~\cite{Badurina:2022ngn}, both the ULDM signal and dominant irreducible backgrounds in future single-photon atom gradiometers will be Gaussian distributed with zero mean.~Hence, the PSD of the expected signal and background will be exponentially distributed. For a signal-plus-background model $\mathcal{M}_\mathrm{S\,+\,B}$, we define the parameter vector $\boldsymbol{\theta} = \{\boldsymbol{\theta}_\mathrm{sig},\,\boldsymbol{\theta}_\mathrm{nuis}\}$, where $\boldsymbol{\theta}_{\mathrm{sig}}$ describes the signal parameters that characterize the ULDM signal contribution, and $\boldsymbol{\theta}_{\mathrm{nuis}}$ describes the background (nuisance parameters). In this case, the appropriate likelihood\footnote{We note that this likelihood differs from the one obtained for the $\mathcal{N} = 1$ case in Ref.~\cite{Badurina:2022ngn} by a factor of $\sqrt{\pi}$, which cancels out in all relevant test statistics used in this work.} to determine the correct upper limits on the couplings and to claim a discovery is given by
\begin{equation}\label{eqn:likelihood}
\mathcal{L}(d \mid \mathcal{M}, \boldsymbol{\theta}) = \prod_{k=1}^{N-1} \frac{1}{\langle S_k(\boldsymbol{\theta}) \rangle} \exp \left[-\frac{S_{\mathrm{data},\,k}}{\langle S_{k}(\boldsymbol{\theta}) \rangle  }\right] \, ,
\end{equation}
where the product is over frequency indices $k \in \{ 1, 2, ..., N-1 \}$, excluding $k = N/2$, and where 
\begin{equation}
\langle S_{k}(\boldsymbol{\theta}) \rangle = \langle S_{\mathrm{DM},\,k} (\boldsymbol{\theta}_\mathrm{sig}) \rangle + \langle S_{\mathrm{Noise},\,k}(\boldsymbol{\theta}_\mathrm{nuis}) \rangle
\end{equation}
is the sum of the expected PSD of the signal and background. Above $\sim 1$~Hz, atom shot noise is expected to dominate the background~\cite{Badurina:2022ngn}, so we will assume that the expected PSD of the noise is frequency-independent. 

\begin{table*}[t]
    \centering
    \begin{tabular}{ c c }
    \toprule
    Time domain & Frequency domain \\
    \hline
    \midrule
    \begin{tabular}{c}
    ULDM coherence time \\ 
    $\tau_c \equiv 2\pi/(m_\phi v_0 ^2)$
    \end{tabular} & 
    \begin{tabular}{c}
    ULDM signal's linewidth \\ $\Delta f_\phi \sim 1/\tau_c$
    \end{tabular} \\ \hline
    \begin{tabular}{c}
    Sampling interval \\ 
    $\Delta t$
    \end{tabular}
    & 
    \begin{tabular}{c}
    Sampling frequency \\
    $f_s = 1/\Delta t$
    \end{tabular} \\
    \hline
    \begin{tabular}{c}
    Integration time \\
    $T_\mathrm{int}$ 
    \end{tabular} 
    & 
    \begin{tabular}{c}
    Experimental frequency resolution \\
    $\Delta f = 1/T_\mathrm{int}$ 
    \end{tabular} \\ 
    \hline
    \begin{tabular}{c}
    Integration time per stack \\
    $T_\mathrm{int}^\mathrm{st} = T_\mathrm{int}/N_\mathrm{stacks}$
    \end{tabular} 
    & 
    \begin{tabular}{c} 
    Experimental stacked frequency resolution \\ 
    $\Delta f^\mathrm{st} = N_\mathrm{stacks}\Delta f$ 
    \end{tabular} \\
    \hline
    \end{tabular}
     \caption{Dictionary of conjugate variables in the time (left column) and frequency (right column) domains pertinent to the analysis of a time-dependent ULDM signal.}
     \label{tab:Dictionary}
\end{table*}

To reduce the number of frequency bins \textit{post-data collection} over which to evaluate the likelihood, which drastically shortens the computational time required to evaluate the test statistic (TS) to claim a discovery or set upper limits without changing their value~\cite{Foster:2017hbq}, it is possible to: \emph{i}) break up the time series into $N_\mathrm{stacks}$ chunks of duration $T_\mathrm{int}/N_\mathrm{stacks}$; \emph{ii}) compute the PSD on each time series, which we label as $S_{\mathrm{data},\,k}^{(l)}$\,; and \emph{iii}) compute the average of these PSD for each frequency bin $k \in \{ 1, ..., N/N_\mathrm{stacks}-1\}$, excluding $k = N/2N_\mathrm{stacks}$. We refer to this averaged quantity as the ``stacked PSD'', which we define mathematically as
\begin{equation}
    \overline{S}_{\mathrm{data},\,k} = \frac{1}{N_\mathrm{stacks}}\sum_{l \, = \, 0}^{N_\mathrm{stacks} \,- \,1} S_{\mathrm{data},\,k}^{(l)} \, .
\end{equation}
The distribution of the sum of $N_\mathrm{stacks}$-independent and identically distributed random variables, each having an exponential distribution with the same mean, is given by the Erlang distribution~\cite{Forbes2010-et}.~Upon a change of variable, the probability distribution function (PDF) of the stacked PSD is then given by the re-scaled Erlang distribution
\begin{equation}\label{eqn:re-scaled_Erlang}
    P\left[\, \overline{S}_{{\mathrm{data}}, k} \right]=\frac{N_\mathrm{stacks}^{N_\mathrm{stacks}}}{\left(N_\mathrm{stacks}-1\right) !} \frac{\left(\, \overline{S}_{\mathrm{data},\,k}\right)^{N_\mathrm{stacks}-1}}{\langle S_k \rangle^{N_\mathrm{stacks}}} \exp \left [-\frac{N_\mathrm{stacks} \, \overline{S}_{\mathrm{data},\,k}} {\langle S_k \rangle} \right ] \,,
\end{equation}
where $\langle S_k \rangle$ is the expectation value of the PSD at the $k^\mathrm{th}$ frequency bin. It then follows that the likelihood\footnote{We note that the likelihood defined in Eq.~\eqref{eqn:stacks_like} differs from the re-scaled Erlang distribution as defined in Eq.~\eqref{eqn:re-scaled_Erlang}.~Indeed, we infer the form of the likelihood by dividing Eq.~\eqref{eqn:re-scaled_Erlang} by a factor of $\left(\,\overline{S}_{\mathrm{data}, k}\right)^{N_\mathrm{stacks}-1} N_\mathrm{stacks}^{N_\mathrm{stacks}}/\left(N_\mathrm{stacks}-1\right) !$. We justify this by noting that this factor cancels in all relevant test statistics, which consist of the differences between logarithms of Eq.~\eqref{eqn:stacks_like}.}  for the stacked data can be defined as
\begin{equation}\label{eqn:stacks_like}
    \mathcal{L}_\mathrm{stacks}(d \mid \mathcal{M}, \boldsymbol{\theta}) = \prod_{k=1}^{N/N_\mathrm{stacks}-1} \frac{1}{\langle S_k(\boldsymbol{\theta}) \rangle^{N_\mathrm{stacks}}} \exp \left [-\frac{N_\mathrm{stacks} \, \overline{S}_{\mathrm{data},\,k}} {\langle S_ k(\boldsymbol{\theta}) \rangle} \right ] \,, 
\end{equation}
where
\begin{equation}
    \langle S_{k}(\boldsymbol{\theta}) \rangle = \langle S_{\mathrm{DM},\,k} (\boldsymbol{\theta}_\mathrm{sig}) \rangle + \langle S_{\mathrm{Noise},\,k}(\boldsymbol{\theta}_\mathrm{nuis}) \rangle \,.
\end{equation}
In light of the reduction in the effective integration time by a factor of $N_\mathrm{stacks}$, the frequency resolution will be decreased by a factor of $N_\mathrm{stacks}$, as shown in Table~\ref{tab:Dictionary}. Thus, in order to resolve the spectral features of a putative ULDM signal with $f_\phi \sim 1$~Hz which is integrated over a time $T_\mathrm{int} = 10^8$\,s, an experimentalist should choose $N_\mathrm{stacks} \lesssim 10^{2}$. 

\subsection{Test statistic for discovery} \label{subsec: ts for discovery}

With these likelihoods, we can define the discovery test statistic, namely
\begin{equation}\label{eq: TS}
    \mathrm{TS}(\boldsymbol{\theta}_\mathrm{sig}) = 2\ln{\frac{\mathcal{L} \big(d \mid \mathcal{M}_{S\,+\,B}, \big\{\widehat{\widehat{\boldsymbol{\theta}}}_\mathrm{nuis}, \, \boldsymbol{\theta}_\mathrm{sig} \big\} \big)}{\mathcal{L}(d \mid \mathcal{M}_B, \{  \widehat{\boldsymbol{\theta}}_\mathrm{nuis}\})}} \, ,
\end{equation}
where $\widehat{\boldsymbol{\theta}}_{\text {nuis}}$ denotes the vector of nuisance parameters that maximise the likelihood in the background-only hypothesis (i.e., the denominator term), and $\widehat{\widehat{\boldsymbol{\theta}}}_{\text {nuis}}$ represents the vector of nuisance parameters that maximise the likelihood in the signal-plus-background hypothesis for a given set of signal parameters. The best fit point in ULDM parameter space is then defined by the vector $\boldsymbol{\widehat{\theta}}_\mathrm{sig}$ that maximises Eq.~\eqref{eq: TS}. Unless otherwise stated, we set $\boldsymbol{\theta}_\mathrm{sig} = (m_\phi,\,d_\phi^2)$.\footnote{Although the signal phase shift is proportional to $d_\phi$, the signal PSD is proportional to $d_\phi^2$. By performing the analysis in the frequency-domain, it is therefore natural to search for a ULDM signal by looking for fluctuations in the amplitude PSD. This justifies $d_\phi^2$ as a parameter of interest in the signal parameter vector $\boldsymbol{\theta}_{\mathrm{sig}}$.}

Since the likelihoods are Gaussian, we define the confidence region (C.R.) with confidence level (C.L.) $\alpha$ in ULDM parameter space as the set $\{\boldsymbol{\theta}_\mathrm{sig}\}$  satisfying
\begin{equation} \label{eq:CR}
\mathrm{TS}_\mathrm{max} - \mathrm{TS}(\boldsymbol{\theta}_\mathrm{sig}) \geq Q_\alpha \, , 
\end{equation}
where $ \mathrm{TS}_\mathrm{max} = \mathrm{TS}(\boldsymbol{\widehat{\theta}}_\mathrm{sig})$ corresponds to the maximum value of the discovery test statistic, and $Q_\alpha$ is a quantile of order $\alpha$ of the $\chi^2$ distribution, and as such depends on the confidence level $\alpha$ and the number of fitted parameters~\cite{StatsBookCowan}. When a ULDM signal is non-zero in multiple frequency bins, the Wald approximation and Wilks' theorem~\cite{Cowan:2010js} are valid. In this case, the C.R. at the 68.3\% and 95.4\% C.L. are associated with $Q_{68.3\%} = 2.3$ and $Q_{95.4\%} \approx 5.99$, respectively.

Obtaining the threshold value of the test statistic to determine the \emph{global} significance of a signal is involved and requires the application of the \emph{look elsewhere effect}. The simplest approach would be to evaluate the test statistic for discovery at fixed masses over a range of independent (i.e., non-overlapping) frequency windows, whose width is given by the expected line-width of the ULDM signal. With this prescription, we can estimate the number of independent frequency windows given a minimum and maximum frequency over which to perform the scan, $f_\mathrm{min}$ and $f_\mathrm{max}$ respectively, as $N_\mathrm{windows} \approx 10^6 \, \ln (f_\mathrm{max}/f_\mathrm{min})$ for $N_\mathrm{windows}\gg 1$.~Hence, the threshold for claiming a discovery can be related to the $p$-value via
\begin{equation}
    \sqrt{\mathrm{TS}_\mathrm{thresh}} = \Phi^{-1} \left( 1-\frac{p}{N_\mathrm{windows}}\right) \, ,
\end{equation}
where $\Phi^{-1}$ is the inverse cumulative distribution function of the
normal distribution. For example, a $5\sigma$ local discovery (i.e. $N_\mathrm{windows} = 1$) would require $p \approx 2.87\times10^{-7}$ and $\sqrt{\mathrm{TS}_\mathrm{thresh}} = 5$. If we scanned between 2~Hz and 22~Hz (i.e. $N_\mathrm{windows} \approx 2.4\times10^6$), a $5\sigma$ global discovery (i.e. $N_\mathrm{windows} = 1$) would instead require $p \approx 2.87\times10^{-7}$ and $\sqrt{\mathrm{TS}_\mathrm{thresh}} \approx 7.32$.\footnote{Alternatively, to avoid performing a scan over independent and non-overlapping frequency windows, which can only be performed at the expense of the frequency resolution, the global $p$-value can be inferred from MC realisations of the background-only hypothesis using the procedure elucidated in Refs.~\cite{Gross:2010qma, Vitells_2011}.}

From these likelihoods it is also possible to define the test statistic for setting upper limits on $d_\phi$ at specific mass values. Since the focus of this work is on understanding how to discover a super-Nyquist signal, we provide a short discussion on this test statistic in Appendix~\ref{app: q for upper lims}.


\section{Discovering a super-Nyquist ULDM signal}\label{sec: high-freq analysis}

In this section, we will explore the approach towards discovering a super-Nyquist ULDM signal in a broadband atom gradiometer.~After introducing the concept of aliasing, we show how a signal oscillating close to or beyond the Nyquist frequency of the experiment can be correctly reconstructed by making use of the framework presented in section~\ref{sec: freq-dom analysis}.

\subsection{An overview of aliasing, Nyquist windows, and folding}\label{subsec: intro aliasing}

The well-motivated high-frequency window to which vertical atom gradiometers would be sensitive lies above $\mathcal{O}(1~\mathrm{Hz})$, which is greater than and comparable to the \textit{Nyquist frequency} $f_\mathrm{Ny} = f_s/2$ of projected experiments. As is well-known in Fourier analysis~\cite{10.5555/227373, 10.5555/294797}, any signal whose spectral content is greater than $f_\mathrm{Ny}$ will be mapped to frequencies between 0 and $f_\mathrm{Ny}$ due to a phenomenon known as \emph{aliasing}. Specifically, it follows from the Nyquist-Shannon theorem \cite{Shannon:1949klj} that a signal oscillating with frequency $f$ is detected at two frequencies between 0 and $f_s$: $f^*_1 = f - \kappa f_s $ and $f^*_2 = (\kappa + 1) f_s - f$, where $\kappa$ is the largest non-negative integer for which $0 < f^*_1 < f_s$. Hence, the spectrum of a signal oscillating at a frequency greater than $f_\mathrm{Ny}$ and aliased to $f_1^*$ will be added to the spectrum of a non-aliased signal oscillating with frequency $f_1^*$. By definition, $f_1^*$ will be measured either in the range $[0, f_\mathrm{Ny}]$ or in the range $[f_\mathrm{Ny},f_s]$, which we refer to as the \textit{first} and \textit{second Nyquist window}, respectively. Hence, if $f_1^*$ is in the first Nyquist window, then $f_2^*$ is in the second Nyquist window, and vice versa. Since $f_1^*$ is related to $f$ by a frequency shift, the original lineshape of the spectrum will be preserved at $f_1^*$; this is to be contrasted with the spectrum identified at $f_2^*$, which is related to $f$ by both a frequency shift and a parity transformation, and so will be a mirror image of the signal's original lineshape. This phenomenon is commonly referred to in the literature as \textit{folding}~\cite{10.5555/227373, 10.5555/294797}. In this sense, the Nyquist frequency acts as an axis of reflection, so that the spectral content measured in the first Nyquist window is a mirror image of the spectral content in the second Nyquist window. All of these principles follow from the symmetries of the Fourier transform, which we review in Appendix~\ref{app: symmetries of the FT}. 

\begin{figure}
    \centering
    \includegraphics[width = \textwidth]{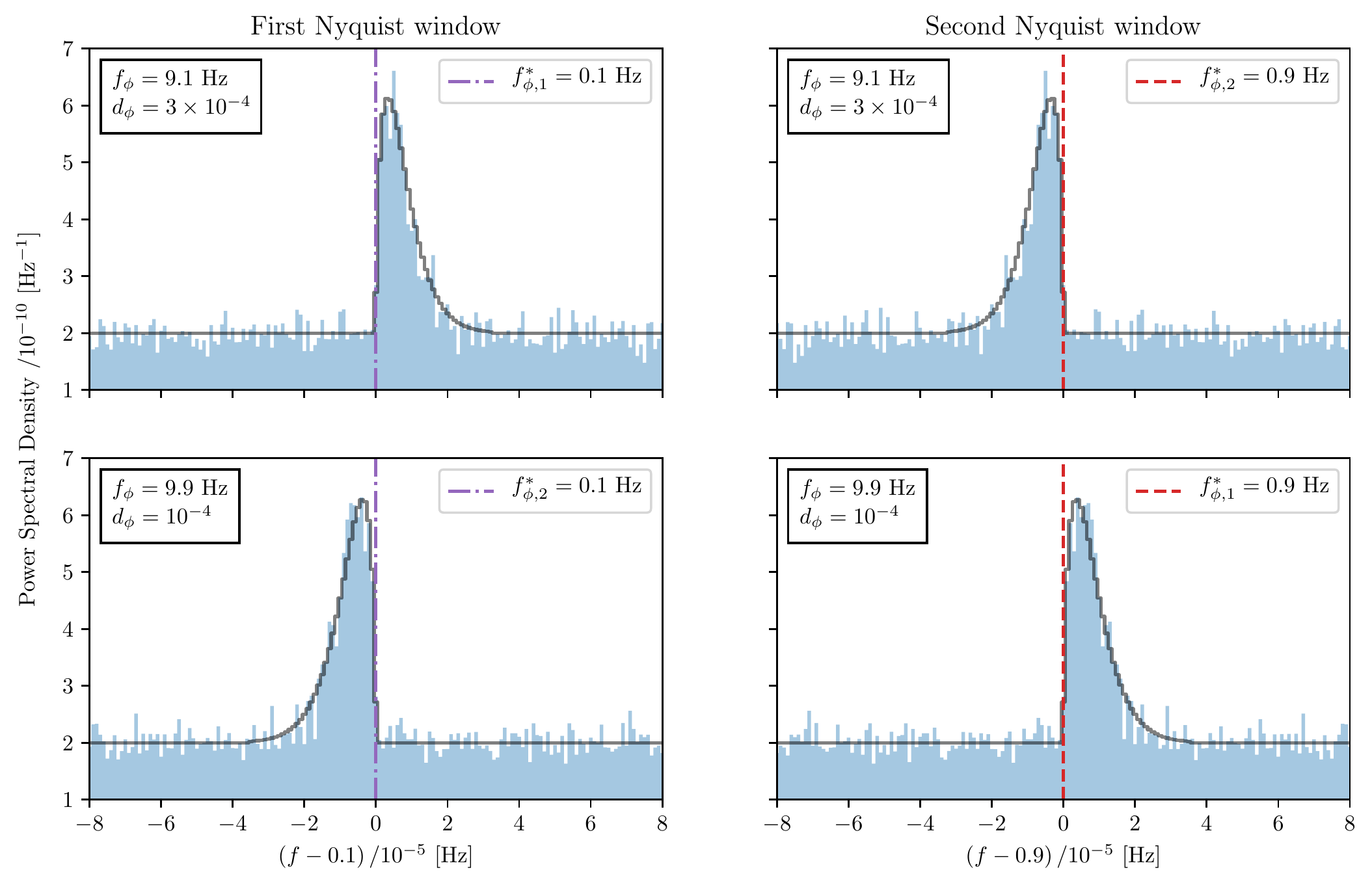} \\
    \caption{Comparison between the expected (black line) and MC-generated (filled bars) PSDs of an injected super-Nyquist ULDM signal with $f_\phi = 9.1$~Hz and $d_\phi = 3 \times 10^{-4}$ (top row), and $f_\phi = 9.9$~Hz and $d_\phi = 10^{-4}$ (bottom row) for the experimental parameters shown in Table~\ref{table: experimental parameters} and $N_\mathrm{stacks} = 100$. In the left column, the aliased signals are mapped to the first Nyquist window, whereas in the right column the signals are mapped to the second Nyquist window. Spectral folding occurs for signals whose aliased frequency is $f_{\phi,2}^*$, and so can be seen in the second Nyquist window for $f_\phi = 9.1$~Hz and in the first Nyquist window for $f_\phi = 9.9$~Hz. For comparison, we also show the aliased ULDM mass at $0.1~(0.9)$~Hz with a purple dash-dotted (red dotted) line in the left (right) panels.}
    \label{fig: MC example}
\end{figure}

To illustrate aliasing in the context of ULDM searches, in the top row of Fig.~\ref{fig: MC example} we show a MC realisation of the broadband signal induced by a scalar ULDM candidate with mass $m_\phi = 2\pi f_\phi = 2 \pi \times 9.1$~Hz and coupling strength $d_\phi = 3 \times 10^{-4}$ that would be measured by a gradiometer which operates with the parameters in Table~\ref{table: experimental parameters} and $N_\mathrm{stacks} = 100$. In this case, $f_{\phi,1}^* = 0.1$~Hz and $f_{\phi,2}^* = 0.9$~Hz; hence, $f_{\phi,1}^*$ and $f_{\phi,2}^*$ are measured in the first and second Nyquist windows, respectively. Since $f_{\phi,1}^*$ is contained in the first Nyquist window, the spectrum identified at $f_{\phi,1}^* = 0.1$ will not be affected by folding. This can be clearly seen in the upper left panel, where the signal rises sharply around the mass of the signal and falls at high frequencies. On the other hand, $f_{\phi,2}^*$, which is contained in the second Nyquist window, will be affected by folding. Indeed, as shown in the upper right panel, the signal now rises sharply at $f_{\phi,2}^* = 0.9$~Hz and falls at low frequencies. For the sake of completeness, in the second row of Fig.~\ref{fig: MC example} we show a MC realisation of the broadband signal induced by a scalar with mass $m_\phi = 2\pi f_\phi = 2 \pi \times 9.9$~Hz and coupling strength $d_\phi = 10^{-4}$ as measured by the same instrument. In this case, $f_{\phi,1}^* = 0.9$~Hz and $f_{\phi,2}^* = 0.1$~Hz, so that the aliased spectrum in the first Nyquist window will now be affected by folding, whilst the aliased spectrum in the second Nyquist window will not. 

\subsection{Disentangling an \textit{aliased} from a \textit{non-aliased} ULDM signal} \label{subsec: alias vs nonalias}

To discover ULDM in the super-Nyquist frequency range it is imperative to be able to distinguish between aliased and non-aliased signals. 
This can only be achieved when a subset of the signal's features is unaffected by aliasing. 
In the context of scalar ULDM searches with broadband atom gradiometers, this set includes the amplitude of the signal and its spectral line-width. 
Owing to aliasing, a ULDM signal with frequency $f_\phi > f_{\rm{Nyq}}$ would be identified at a smaller frequency between zero and $f_s$, but crucially, 
would inherit its original frequency spread and amplitude. This is because the signal amplitude depends on both the coupling strength and the {\it ULDM mass}, 
while the spectral width depends on the properties of the dark matter's speed distribution and the {\it ULDM mass}.

\begin{figure}[t]
    \centering
    \includegraphics[width = \textwidth]{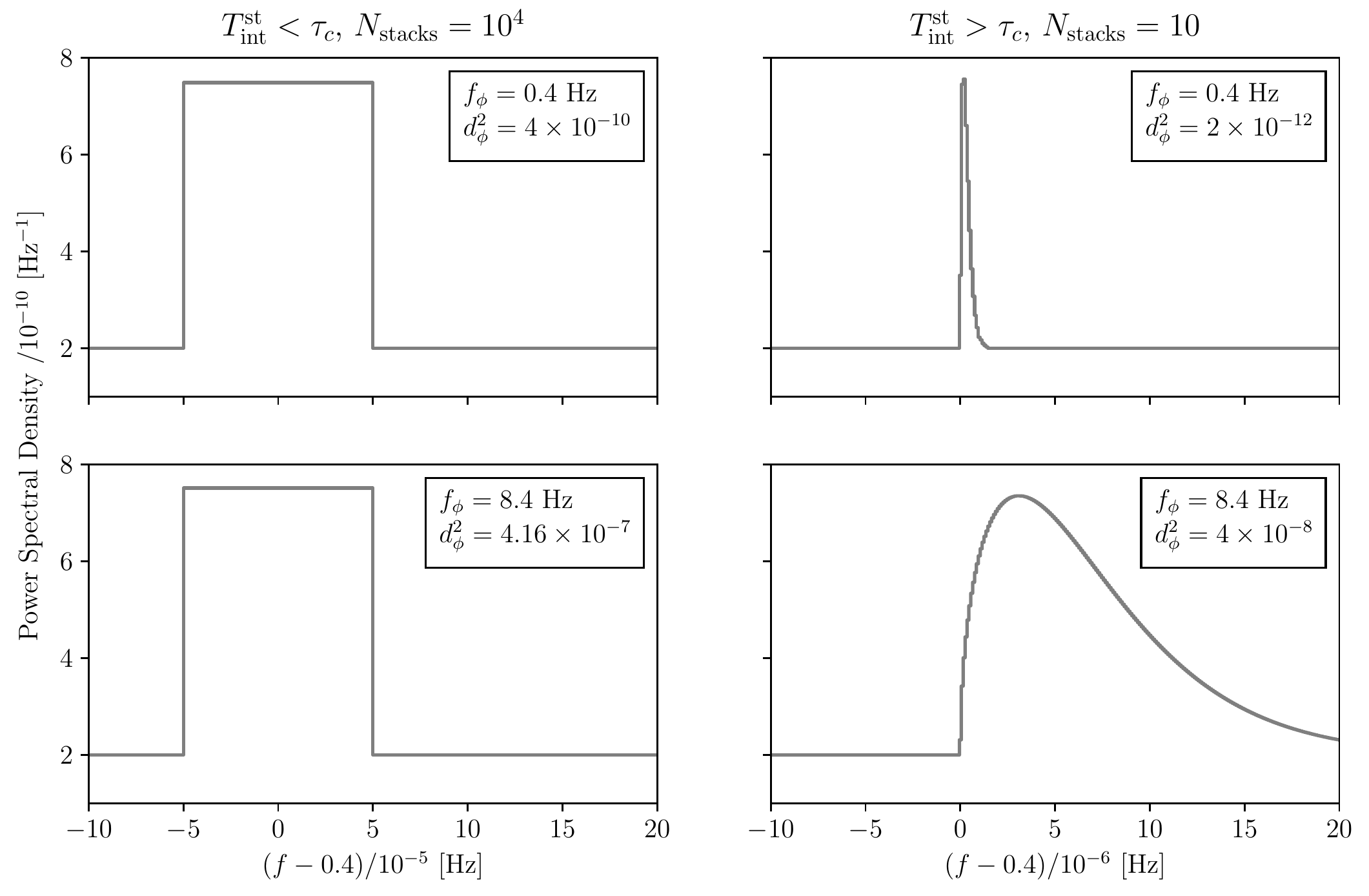}
    \caption{Expected power spectrum densities of aliased and non-aliased ULDM signals which satisfy either $T_\mathrm{int}^\mathrm{st} < \tau_c$ (left column) or $T_\mathrm{int}^\mathrm{st}  > \tau_c$ (right column). In the top row we show the spectra of a ULDM signal with $f_\phi = 0.4$~Hz, which is not affected by aliasing in light of the chosen sampling frequency (see Table~\ref{table: experimental parameters}), for different coupling strengths. In the second row, we show the spectra of a ULDM signal with $f_\phi = 8.4$~Hz, which is subject to aliasing, for different coupling strengths. In the regime of short stacked integration time, the aliased and non-aliased signals that we show are identical; in the opposite regime, the aliased and non-aliased signal are characterised by a marked difference in spectral broadening.}
    \label{fig: aliased example}
\end{figure}

Prima facie, it would then seem that an aliased ULDM signal could always be correctly disentangled from a non-aliased one. This statement, however, is not correct. Indeed, because of the amplitude's degeneracy with coupling strength, aliased and non-aliased signals cannot be disentangled when their spectral content is contained within a single frequency bin, which occurs when the stacked integration time $T^\mathrm{st}_\mathrm{int} = T_\mathrm{int}/N_\mathrm{stacks}$ exceeds the ULDM's coherence time. Therefore, even if $T_\mathrm{int} > \tau_c$ in the frequency range of interest, a post-data collection choice of stacking could weaken the ability to distinguish between aliased and non-aliased signals.  
To illustrate this point, let's consider an experiment operating with the parameters shown in Table~\ref{table: experimental parameters}, and  hunting for signals satisfying $T_\mathrm{int} > \tau_c$. In the left column of Fig.~\ref{fig: aliased example} we show the expected power spectrum density of a non-aliased signal at $f_\phi = 0.4$~Hz with $d_\phi^2 = 4\times 10^{-10}$, and the expected power spectrum density of an aliased signal at $f_\phi = 8.4$~Hz with $d_\phi^2 = 4.16 \times 10^{-7}$, assuming aggressive stacking ($N_\mathrm{stacks} = 10^{4}$). In light of the sampling frequency, both signals are identified at 0.1~Hz; since the stacked integration time $T_\mathrm{int}^\mathrm{st} = 10^4$~s satisfies the condition $T_\mathrm{int}^\mathrm{st} < \tau_c$ for both ULDM masses, both signals are contained in a single frequency bin. Therefore, despite the vastly different phenomenological parameters, both signals appear identical and hence cannot be distinguished. 

Instead, the correct statement is the following: aliased and non-aliased signals can only be disentangled when their spectral content is spread over multiple frequency bins. 
Indeed, in this regime, aliased and non-aliased signals that are mapped to the same frequency between zero and $f_s$, and have identical maximum PSD amplitude, can still be distinguished because of the spectral line-width's linear dependence on the ULDM mass (or equivalently, $f_\phi$). To illustrate this point, we consider the usual experiment operating with the parameters shown in Table~\ref{table: experimental parameters}, but in the case of limited stacking ($N_\mathrm{stacks} = 10$). For this choice of stacking and integration time, the non-aliased signal at $f_\phi = 0.4$~Hz with $d_\phi^2 = 2 \times 10^{-12}$ and the aliased signal at $f_\phi = 8.4$~Hz with $d_\phi^2 = 4 \times 10^{-8}$ satisfy the condition $T_\mathrm{int}^\mathrm{st} > \tau_c$. In the right column of Fig.~\ref{fig: aliased example} we show the expected PSD of these two signals. In each case, the signals are mapped to $f_{\phi, 1}^* = 0.4$~Hz but exhibit significantly different spectral broadening. Since the spectral width of a ULDM signal scales linearly with $f_\phi$, the PSD of the $f_\phi = 8.4$~Hz signal is the broadest. Hence, despite their identical maximum amplitude, these two signals can be easily distinguished.

\begin{figure}[t]
    \centering
    \includegraphics[width = 0.48\textwidth]{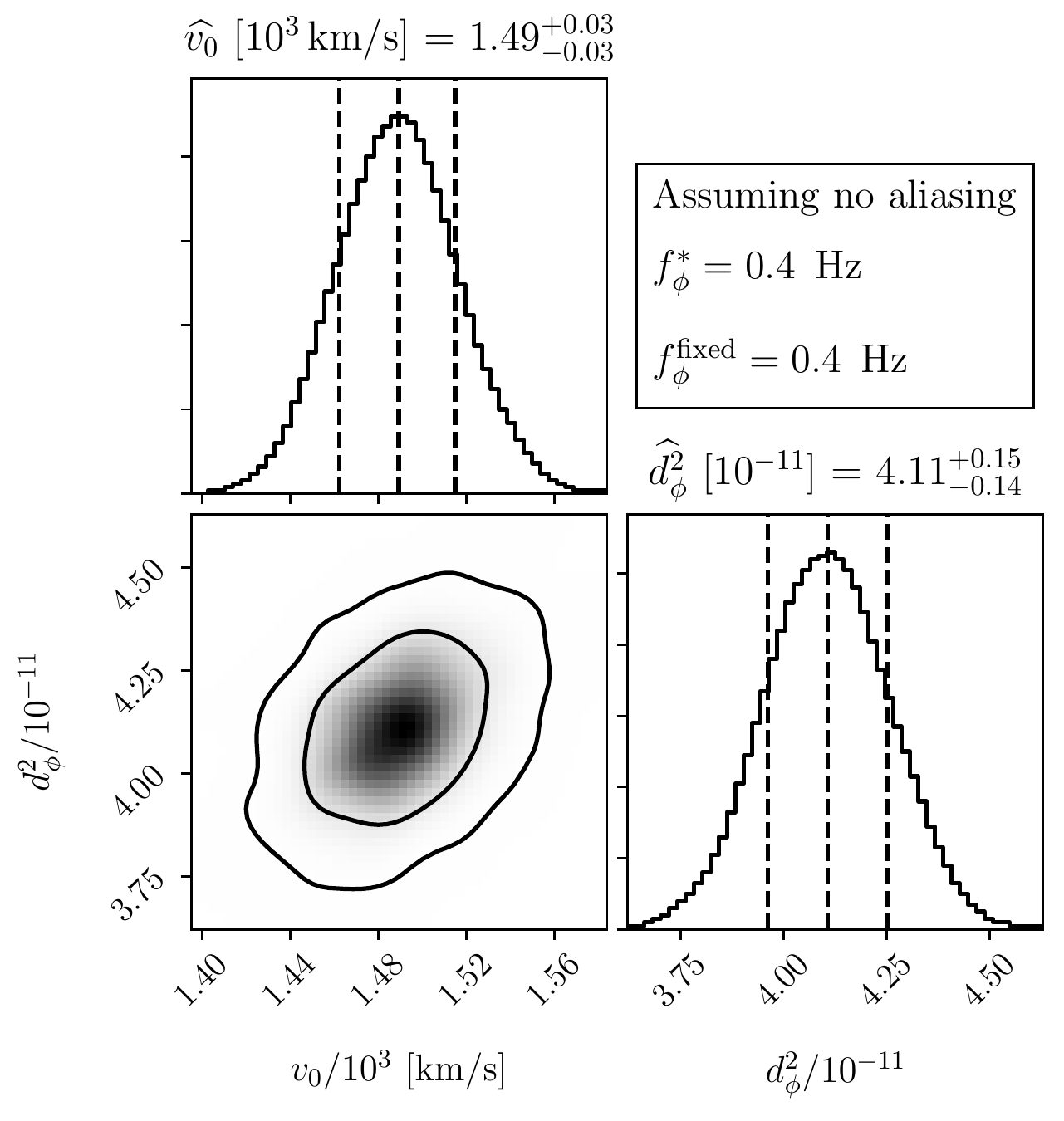}
    \includegraphics[width = 0.49\textwidth]{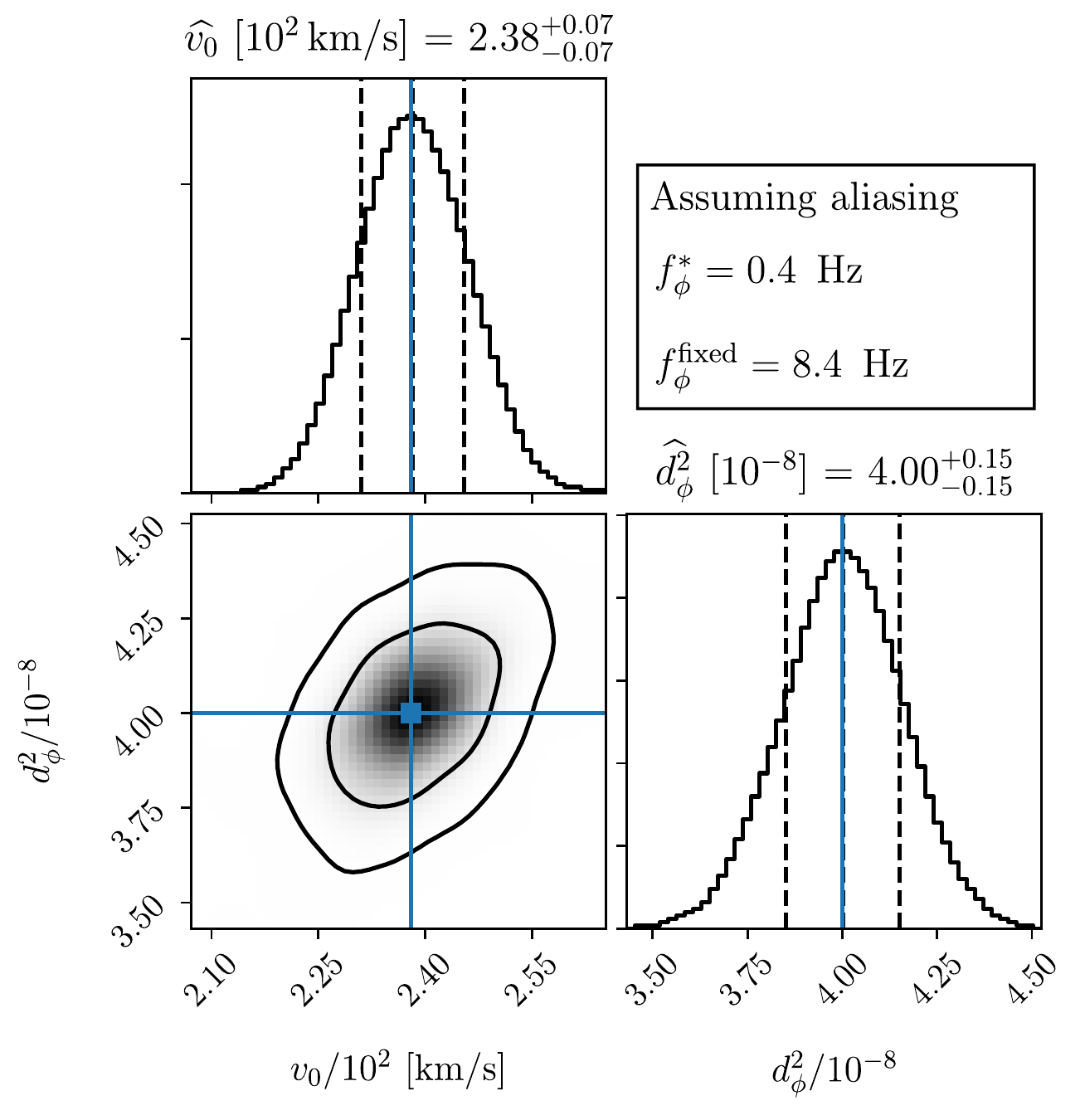}
    \caption{Comparison of the posterior distributions of $d_\phi^2$ and $v_0$ for an injected signal with $f_\phi = 8.4$~Hz and $d_\phi^{2} = 4 \times 10^{-8}$ at two fixed masses: $f_\phi^\mathrm{fixed} = 0.4$~Hz, which assumes no aliasing (left panel), and $f_\phi^\mathrm{fixed} = 8.4$~Hz, which assumes aliasing (right panel). The $1\sigma$ and $2\sigma$ credible regions are shown with solid lines. The injected $(v_0,\,d_\phi^2)$ value is shown in the 2D plane of the right panel by a blue square; it is not visible in the left panel due to different axis scales. Here, we have taken the data set to be equal to the mean predictions of the model under consideration and neglecting statistical fluctuations, i.e. the Asimov approach.}
    \label{fig: contours for 0.1 Hz vs 9.1 Hz}
\end{figure}

If we did not take aliasing into account, an aliased signal satisfying the condition $T_\mathrm{int}^\mathrm{st} > \tau_c$ would be confused for a non-aliased one with different coupling strength and much larger (and unphysical) velocity dispersion, which we remind the reader is given by $v_0/\sqrt{2}$.~We show this in Fig.~\ref{fig: contours for 0.1 Hz vs 9.1 Hz}, where we provide a comparison of the reconstructed coupling and $v_0$ of an injected ULDM signal using the discovery test statistic defined in Eq.~\eqref{eq: TS} at fixed mass. In particular, we analyse the Asimov data set\footnote{The Asimov data set corresponds to taking the data to be equal to the
mean predictions of the model under consideration, and neglecting statistical fluctuations~\cite{Cowan:2010js}.} of a signal with $f_\phi = 8.4$~Hz, $d_\phi^2 = 4\times10^{-8}$ and $v_0 = 238$~km/s at two masses: $f_\phi^\mathrm{fixed} = 0.4$~Hz, which corresponds to the alias of the injected signal's angular frequency and assumes no aliasing, and $f_\phi^\mathrm{fixed} = 8.4$~Hz, which corresponds to the angular frequency of the injected signal and assumes aliasing. In the former case, which is displayed in the left panel, we reconstruct a signal with $\widehat{v_0} = 1490^{+30}_{-30}$~km/s and $\widehat{d_\phi^2} = 4.11^{+0.15}_{-0.14}\times 10^{-11}$. This is to be contrasted with the injected signal parameters, which instead are in agreement with the fitted parameters for $f_\phi^\mathrm{fixed} = 8.4$~Hz: $\widehat{v_0} = 238^{+7}_{-7}$~km/s and $\widehat{d_\phi^2} = 4.00^{+0.15}_{-0.15} \times 10^{-8}$. In particular, in the former case, the inferred best-fit value of $v_0$ is an order of magnitude larger than the prediction of the SHM in Eq.~\eqref{eqn:SHM_dist}, thereby implying that the signal at 0.4~Hz cannot be confidently attributed to ULDM. 

In summary, in light of the difficulty in disentangling a non-aliased signal from an aliased one, we conclude that:~\textit{i}) $N_\mathrm{stacks}$ should be chosen post-data collection so that the effective integration time (i.e., the duration of each stacked time series, $T_\mathrm{int}^\mathrm{st}$) should be larger than the largest coherence time that the experimentalist wishes to probe; and \textit{ii}) the speed parameters should be constrained by the predictions of the SHM, so that the spectral breadth of the signal depends exclusively on the ULDM mass. 

\subsection{Disentangling a \textit{folded} from a \textit{non-folded} ULDM signal}\label{subsec: folding}

In light of the Nyquist-Shannon theorem, the spectral content of the first and second Nyquist windows is identical under a parity transformation. Specifically, the spectrum measured at $f^*_{\phi, 1} = f_\phi - \kappa f_s $ is identical to the one measured at $f^*_{\phi, 2} = (\kappa + 1) f_s - f_\phi$, where $\kappa$ is the largest non-negative integer for which $0 < f^*_{\phi, 1} < f_s$. However, the aliased copy identified at $f^*_{\phi, 1}$ will preserve the signal spectrum's original orientation, while the aliased copy identified at $f^*_{\phi, 2}$ will be its mirror image. 

The degree to which the excesses at $\sim f^*_{\phi, 1}$ and $\sim f^*_{\phi, 2}$ can be correctly identified with the folded and non-folded alias, respectively, of a super-Nyquist ULDM candidate depends on the resolution of the characteristic lineshape of the ULDM signal, which in turn depends on the stacked integration time. To measure this, we introduce the \textit{folding discriminant}, which quantifies the degree to which an excess is attributed to a folded or a non-folded alias. The definition of this measure relies on the properties of an injected ULDM signal $f_\phi'$. We choose to have a non-folded alias in the first Nyquist window, and therefore a folded alias in the second window. The folding discriminant is then defined as the difference between the maximised test statistic for discovery assuming that: the first Nyquist window features a non-folded alias of a signal with frequency $\sim f_\phi'$; and the second Nyquist window contains a non-folded alias of a different signal with frequency $\sim f_\phi''$. Mathematically, this is equivalent~to
\begin{equation}\label{eq: folding discriminant}
Q^\mathrm{FD} = \max_{f_\phi \, \in \, \mathcal{N}_{f_\phi'}} \mathrm{TS}_{1}\left (\left \{f_\phi, \widehat{d_\phi^2}\right\}\right) - \max_{f_\phi \, \in \, \mathcal{N}_{f_\phi''}} \mathrm{TS}_{2}\left (\left \{f_\phi, \widehat{\widehat{d_\phi^2}}\right \}\right ) \, ,
\end{equation}
where $\mathcal{N}_{f_\phi}$ is defined as the neighbourhood around $f_\phi$. 
Since the injected signal contains a non-folded alias in the first Nyquist window only, in Eq.~\eqref{eq: folding discriminant} the second term will be bounded above by the first term. The difference between these test statistics is then analogous to the definition of quantile $Q_\alpha$ of the $\chi^2$ distribution which was used in the definition of confidence regions in ULDM parameter space (cf. Eq.~\eqref{eq:CR}). Hence, the larger the difference between these two test statistics (i.e. the larger the value of the folding discriminant), the larger the discriminating power between folded and non-folded signals.

\begin{figure}
    \centering
    \includegraphics[width = 0.98\textwidth]{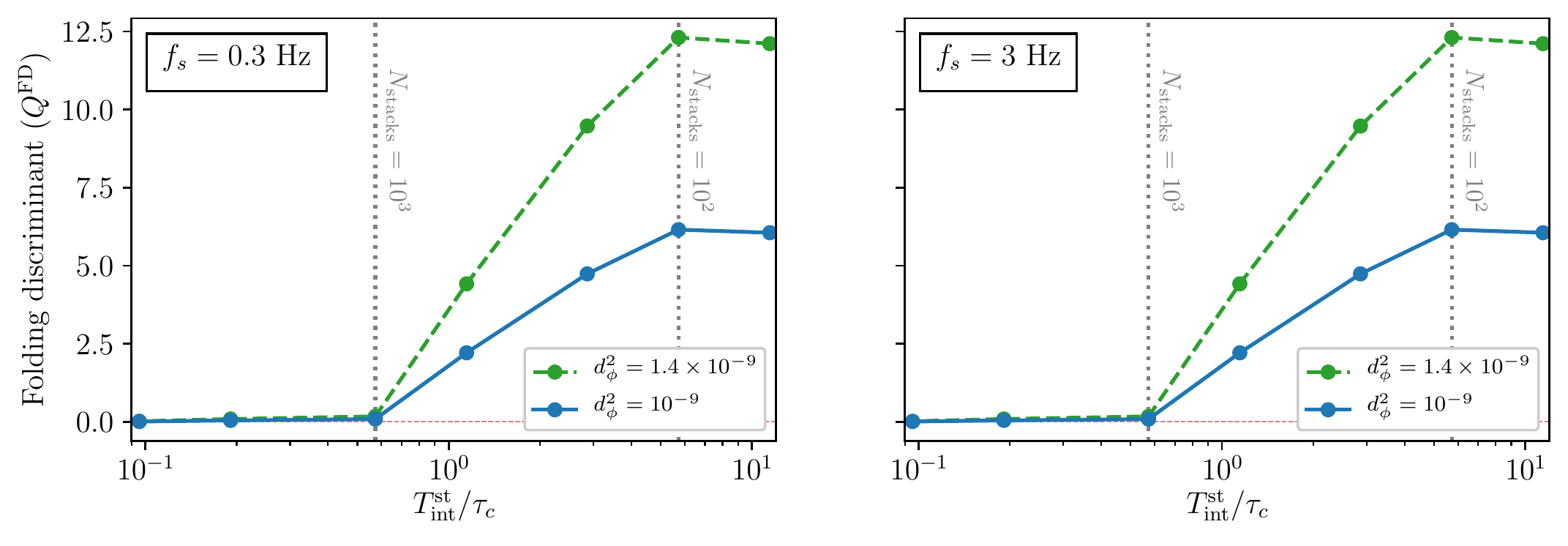}
    \caption{Significance comparison between neighbouring folded and non-folded signals as a function of the ratio of the stacked integration time $T_\mathrm{int}^\mathrm{st}$ to the ULDM coherence time $\tau_c$ for $f_\phi = 9.1$~Hz. The comparison is quantified via the folding discriminant, as defined in the body of the paper, which we evaluate on two Asimov datasets containing a ULDM signal with: $f_\phi = 9.1$~Hz and $d_\phi^2 = 10^{-9}$ (solid blue), and $f_\phi = 9.1$~Hz and $d_\phi^2 = 1.4 \times 10^{-9}$ (dashed green). We assume the experimental parameters stated in Table~\ref{table: experimental parameters} with the exception of the sampling frequency, which we set to 0.3~Hz in the left panel and 3~Hz in the right panel. The folding discriminant is saturated by signals satisfying $T_\mathrm{int}/\tau_c \gtrsim 5 $. The similarity between the panels implies that the ability to disentangle folded and non-folded signals is independent of $f_s$.} 
    \label{fig:comparison of TS for folded and non-folded sig}
\end{figure}

In Fig.~\ref{fig:comparison of TS for folded and non-folded sig}, we make use of the folding discriminant on two Asimov data sets containing an injected signal with $\approx 5\sigma$ local significance ($f_\phi = 9.1$~Hz and $d_\phi^2 = 10^{-9}$) and an injected signal with $\approx 5\sigma$ global significance ($f_\phi = 9.1$~Hz and $d_\phi^2 = 1.4 \times 10^{-9}$).\footnote{From section~\ref{subsec: ts for discovery}, a $5\sigma$ globally significant signal implies a value of $\mathrm{TS}_\mathrm{thresh}$ that is approximately twice the value of $\mathrm{TS}_\mathrm{thresh}$ for a $5\sigma$ locally significant discovery. Since $\mathrm{TS}\propto d_\phi^4$, a $5\sigma$ locally significant signal becomes $5\sigma$ globally significant if $d_\phi^2$ increase by a factor of $\sqrt{2}$.} In both cases, we fix $T_\mathrm{int}$ (i.e. the data has been already taken by the experimentalist) but allow for $N_\mathrm{stacks}$ to be tuned, which implies that we are effectively scanning over different values of $T_\mathrm{int}^\mathrm{st}$. We further choose the experimental parameters stated in Table~\ref{table: experimental parameters} and two different values of the sampling frequency: $0.3$~Hz and 3~Hz. For this choice of sampling frequencies, the alias of the injected signal identified in the first Nyquist window will not be affected by folding. The folded alias of the injected signal will instead be identified at 0.2~Hz and 2.9~Hz for $f_s = 0.3$~Hz and $f_s = 3$~Hz, respectively. For both sampling frequencies, we assume that the aliases in the second Nyquist window are attributable to ULDM fields with $f_\phi = 8.9$~Hz, for which the aliases in the second Nyquist window would not be folded. 

As shown in both panels of Fig.~\ref{fig:comparison of TS for folded and non-folded sig}, the folding discriminant is approximately zero and constant for $T_\mathrm{int}/\tau_c \gtrsim 1$, which corresponds to $N_\mathrm{stacks} \lesssim 10^3$ for $T_\mathrm{int}=10^8$~s, independently of the significance of the injected signal. This follows from the fact that the folded and non-folded signal spectra will each be contained within a single bin, which implies that the two signals will be identical (up to a difference in spectral amplitude); hence, the analysis is unable to discriminate between the folded signal at $f_\phi \approx 8.9$~Hz and the non-folded signal at $f_\phi \approx 9.1$~Hz. Furthermore, since increasing $N_\mathrm{stacks}$ (i.e. decreasing $T_\mathrm{int}^\mathrm{st}$) does not improve the resolution of the signal's spectral content, the ability to distinguish between these two signals is independent of $N_\mathrm{stacks}$.

For $T_\mathrm{int}/\tau_c \gtrsim 5$, the folding discriminant is maximal and constant. This can be understood as follows: the limit $T_\mathrm{int} > \tau_c$ implies that the signal's spectral content is well-resolved. In this case, the folded and non-folded aliased copies of the true signal will be characterised by well-resolved line-shapes with opposite parity; hence, the spectra at $f_{\phi, 1}^*$ and $f_{\phi, 2}^*$ can be correctly identified with the non-folded and folded aliases, respectively. Since we have assumed no DM substructure, increasing $T_\mathrm{int}^\mathrm{st}$ further (i.e. choosing $N_\mathrm{stacks} \lesssim 10^2$ for $T_\mathrm{int} = 10^8$~s) does not improve the resolution of the signal's characteristic lineshape; hence the ability to distinguish between these two signals is independent of $N_\mathrm{stacks}$. In both panels, however, the folding discriminant is largest for the signal that has the highest significance (i.e. the signal with the largest coupling), which follows from the definition of the folding discriminant and from the scaling of Eq.~\eqref{eq: TS} with $d_   \phi^2$. This can also be understood as follows: for large couplings, the features of the expected signal's lineshape are more pronounced; hence, the folded and non-folded aliases of ULDM signals with large couplings can be more readily distinguished.

Finally, for our choice of sampling frequencies, $f_{\phi, 1}^*$ and $f_{\phi, 2}^*$ are not affected by $f_s$; hence, the degree to which the likelihood can distinguish between folded and non-folded signals is largely insensitive to the sampling frequency, which explains the substantial similarity between the left and right panels of Fig.~\ref{fig:comparison of TS for folded and non-folded sig}.

In summary, we conclude that: \textit{i}) the degree to which folded and non-folded signals can be distinguished depends on the ability to resolve the signal's characteristic lineshape; \textit{ii)} folded and non-folded signals satisfying $T_\mathrm{int}^\mathrm{st}/\tau_c \gtrsim 5$ have well-resolved lineshapes with opposite parity and so can be readily disentangled, whilst those satisfying $T_\mathrm{int}^\mathrm{st}/\tau_c \lesssim 1$ have unresolved lineshapes and so cannot be correctly reconstructed; and \textit{iv}) the degree to which folded and non-folded signals can be distinguished is independent of the sampling frequency~$f_s$.

\subsection{Disentangling neighbouring \textit{aliased} and \textit{equally-folded} signals} \label{subsec: distinguishing aliased sigs}

From the Nyquist-Shannon theorem, two ULDM signals whose frequencies differ by integer multiples of the sampling frequency are both identified at the same frequency between zero and $f_s$ - i.e. two ULDM signals with $f_\phi' = \kappa f_s + f_\phi^*$ and $f_\phi'' = \kappa' f_s + f_\phi^*$, for integers $\kappa$ and $\kappa'$, would be imaged at $f_\phi^*$. Hence, different from the case of folded and non-folded signals, the spectra of neighbouring aliased signals will have the same parity (i.e. they will exhibit the same degree of folding). Since $\Delta f_\phi \propto f_\phi$, such signals may be differentiated post-data-taking by leveraging exclusively on the spectral linewidth's dependence on the DM mass.

The degree to which such signals can be disentangled largely depends on the sampling frequency of the experiment. In particular, for high-frequency ULDM signals, the larger the sampling frequency, the larger the frequency difference between super-Nyquist signals which are consistent with the same alias; the larger the frequency splitting between neighbouring signals, the larger the linewidth difference between neighbouring signals, and so the greater the incompatibility between such ULDM candidates. To measure this, we introduce the \textit{aliasing discriminant}.
In analogy to the folding discriminant defined in section~\ref{subsec: folding}, the aliasing discriminant is computed using the discovery test statistic. Additionally, the definition of this measure also relies on the properties of an injected ULDM signal $f_\phi$, which we choose to have a non-folded alias in the first Nyquist window. However, different from the folding discriminant, this object is defined as the difference between the test statistic for discovery assuming that: the first Nyquist window contains a non-folded alias of a signal with frequency $\sim f_\phi'$; and the first Nyquist window contains a non-folded alias of a signal with frequency $f_\phi'' = f_\phi' + \kappa f_s$. Mathematically, this is equivalent to
\begin{equation}\label{eq: aliasing discriminant}
Q^\mathrm{AD} = \max_{f_\phi \, \in \, \mathcal{N}_{f_\phi'}} \mathrm{TS}_{1}\left(\left\{f_\phi, \widehat{d_\phi^2}\right\}\right) - \max_{f_\phi \, \in \, \mathcal{N}_{f_\phi''}} \mathrm{TS}_{1}\left(\left\{f_\phi, \widehat{\widehat{d_\phi^2}}\right\}\right) \, .
\end{equation}
Since the injected signal is at $f_\phi$, in Eq.~\eqref{eq: aliasing discriminant} the second term will be bounded above by the first term. Hence, as for the folding discriminant, this object is analogous to the definition of the quantile $Q_\alpha$. Therefore, the larger the difference between these two test statistics (i.e.\ the larger the value of the aliasing discriminant), the larger the discriminating power between super-Nyquist signals with different masses.

\begin{figure}
    \centering
    \includegraphics[width = 0.98\textwidth]{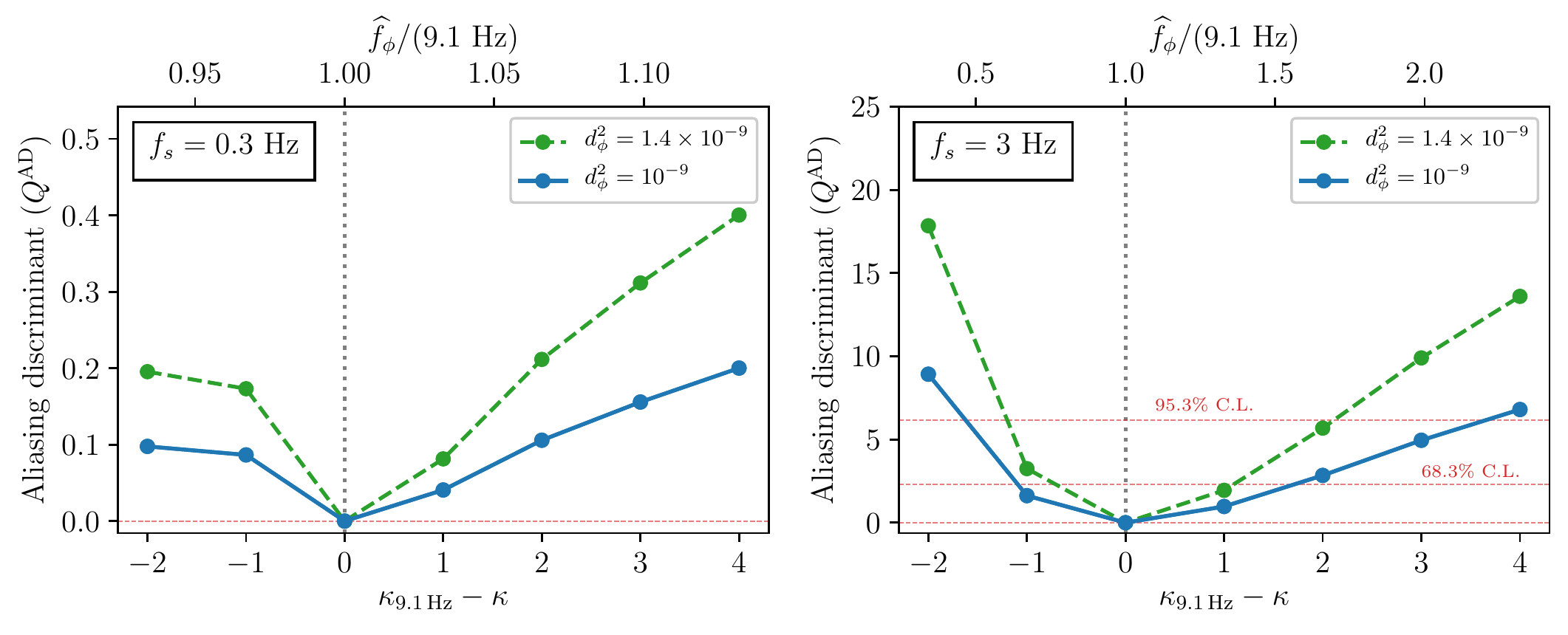}
    \caption{Significance comparison between neighbouring equally-folded signals as a function of integer multiples of the first Nyquist window (lower axis), or equivalently as a function of the frequency (upper axis).
    The comparison is quantified via the aliasing discriminant, as defined in the body of the paper, which we evaluate on two Asimov datasets containing a ULDM signal with: $f_\phi = 9.1$~Hz and $d_\phi^2 = 10^{-9}$ (solid blue) and $f_\phi = 9.1$~Hz and $d_\phi^2 = 1.4 \times 10^{-9}$ (dashed green). We assume the experimental parameters stated in Table~\ref{table: experimental parameters} with the exception of the sampling frequency, which we set to 0.3~Hz in the left panel and 3~Hz in the right panel. The dashed horizontal lines define the quantiles associated with the 68.3\% and 95.4\% C.L. The aliasing discriminant increases away from the injected signal; as the right panel shows, this object is largest for signals whose frequency is much larger or smaller than the injected one.}
    \label{fig:comparison of TS for equally folded sig}
\end{figure}

By making use of the aliasing discriminant, which we evaluate on two Asimov data sets containing an injected signal with $\approx 5\sigma$ local significance ($f_\phi = 9.1$~Hz and $d_\phi^2 = 10^{-9}$) and an injected signal with $\approx 5\sigma$ global significance ($f_\phi = 9.1$~Hz and $d_\phi^2 = 1.4 \times 10^{-9}$), in Fig.~\ref{fig:comparison of TS for equally folded sig} we illustrate the likelihood's ability to distinguish between neighbouring aliased and equally-folded signals as a function of $\kappa$.  Similarly to Fig.~\ref{fig:comparison of TS for folded and non-folded sig}, we focus on two different sampling frequencies: 0.3~Hz, which we display in the left panel, and 3~Hz, which we display in the right panel. Here, however, we set $N_\mathrm{stacks} = 100$. The aliasing discriminant is then computed over ULDM masses contained in integer multiples of the first Nyquist window, which we define as $\kappa$, closest to the one containing 9.1~Hz, which we define as $\kappa_{9.1~\mathrm{Hz}}$, respectively. For illustrative purposes, we restrict ourselves to $28 \leq \kappa \leq 34$ for $f_s = 0.3$~Hz, and $1 \leq \kappa \leq 7$ for $f_s = 3$~Hz. The multiples of the first Nyquist window containing 9.1~Hz for $f_s = 0.3$~Hz and $f_s = 3$~Hz are then $\kappa_{9.1~\mathrm{Hz}} = 30$ and $\kappa_{9.1~\mathrm{Hz}} = 3$, respectively. 

In both panels, the aliasing discriminant is zero and at its minimum for $\kappa = \kappa_{9.1~\mathrm{Hz}}$,  which implies that the reconstructed signal is favoured to be in the correct multiple of the Nyquist window. Furthermore, the aliasing discriminant increases with $|\kappa-\kappa_{9.1~\mathrm{Hz}}|$, which implies that the ability to distinguish between neighbouring and equally-folded signals increases with the difference between the reconstructed signal's expected linewidth. Furthermore, as for the folding discriminant, neighbouring aliased signals with larger couplings (and larger significance) will be more readily distinguished, as shown by the higher values of the aliasing discriminant away from $\kappa_{9.1~\mathrm{Hz}}$. 

The degree to which equally-folded signals in neighbouring multiples of the first Nyquist window are less significant depends on the sampling frequency. For $f_s = 0.3$~Hz, the ratio of $\widehat{f_\phi}$ to 9.1~Hz in neighbouring windows is approximately unity, which implies that the spectral content of these signals is comparable with the injected one; hence, the aliasing discriminant is approximately zero, i.e. neighbouring signals cannot be easily disentangled. For $f_s = 3$~Hz, however, $0.34 \lesssim \widehat{f_\phi}/ (9.1~\mathrm{Hz})\lesssim 2.4$, which implies that the spectral content of these signals is much wider or narrower than that of the injected signal; hence, the aliasing discriminant increases rapidly, i.e. neighbouring aliased signals will be more robustly disfavoured. In particular, we note that reconstructed signals that are at least twice or at most half as wide as the injected signal in the frequency domain would not be contained within the 95.4\%~C.L. of the injected signal. 

From this last observation, we can infer a condition on $f_\phi$ and $f_s$ to maximise the likelihood's ability to distinguish between neighbouring aliased and equally-folded signals. As shown in Fig.~\ref{fig:comparison of TS for equally folded sig}, the aliasing discriminant of signals whose linewidth is at least twice or at most half as wide as that of the injected signal are not contained within the 95.4\% C.L. quantile of the global maximum. Hence, signals satisfying $(f_\phi+ f_s)/f_\phi \gtrsim 2$ and $(f_\phi - f_s)/f_\phi \lesssim 1/2$ will not be in good agreement with the injected signal. Combining these two conditions, we find $f_s/f_\phi \gtrsim 3/5$. 

In summary, we conclude that:~\textit{i}) the degree to which neighbouring equally-folded aliased signals can be disentangled improves with the sampling frequency; and \textit{ii}) in order to contain the $95.4\%$ C.L. confidence region of a $\approx 5\sigma$ globally significant discovery in a single Nyquist window, it will be necessary to choose a value of the sampling frequency that satisfies the condition $f_s/f_\phi \gtrsim 3/5$, where $f_\phi$ is the largest ULDM mass to be considered in the scan; equivalently, for a given choice of $f_s$, which is stipulated before the start of the measurement campaign, $f_\phi = 3\, f_\phi/5$ would be the largest ULDM mass that could be unambiguously reconstructed by our analysis.

\subsection{Discovering \textit{distorted} signals} \label{subsec: spectral distortions}

Another defining feature of aliased signals is distortion due to folding. This arises when the spectral content of a signal exceeds the frequency range of a given Nyquist window; in light of the symmetries of the Fourier transform (see Appendix~\ref{app: symmetries of the FT}), the power spectrum that leaks into the neighbouring Nyquist window is reflected back to the original Nyquist window, and added to the PSD that was initially aliased to this Nyquist window. In the case of ULDM searches, this phenomenon would occur for signals satisfying $f_\phi^* + f_\phi \, v_\mathrm{esc}^2/2 > f_\mathrm{Ny}$, when $f_\phi^*$ is contained within the first Nyquist window, or $f_\phi^* + f_\phi \, v_\mathrm{esc}^2/2 > f_s$, when $f_\phi^*$ is contained within the second Nyquist window. 

All ULDM signals satisfying $f_\phi \gtrsim 10^6 \, f_\mathrm{Ny}$ would be affected by distortions due to folding, independently of $f_\phi^*$. This is because the spectral width of these very high-frequency super-Nyquist signals exceeds the size of a single Nyquist window. Here, we will not consider such signals: this part of parameter space is already competitively probed by complementary probes for typical values of $f_s$, and so would not be a primary target of ULDM searches with broadband atom gradiometers. 

\begin{figure}[t]
    \centering
    \includegraphics[width=0.99\textwidth]{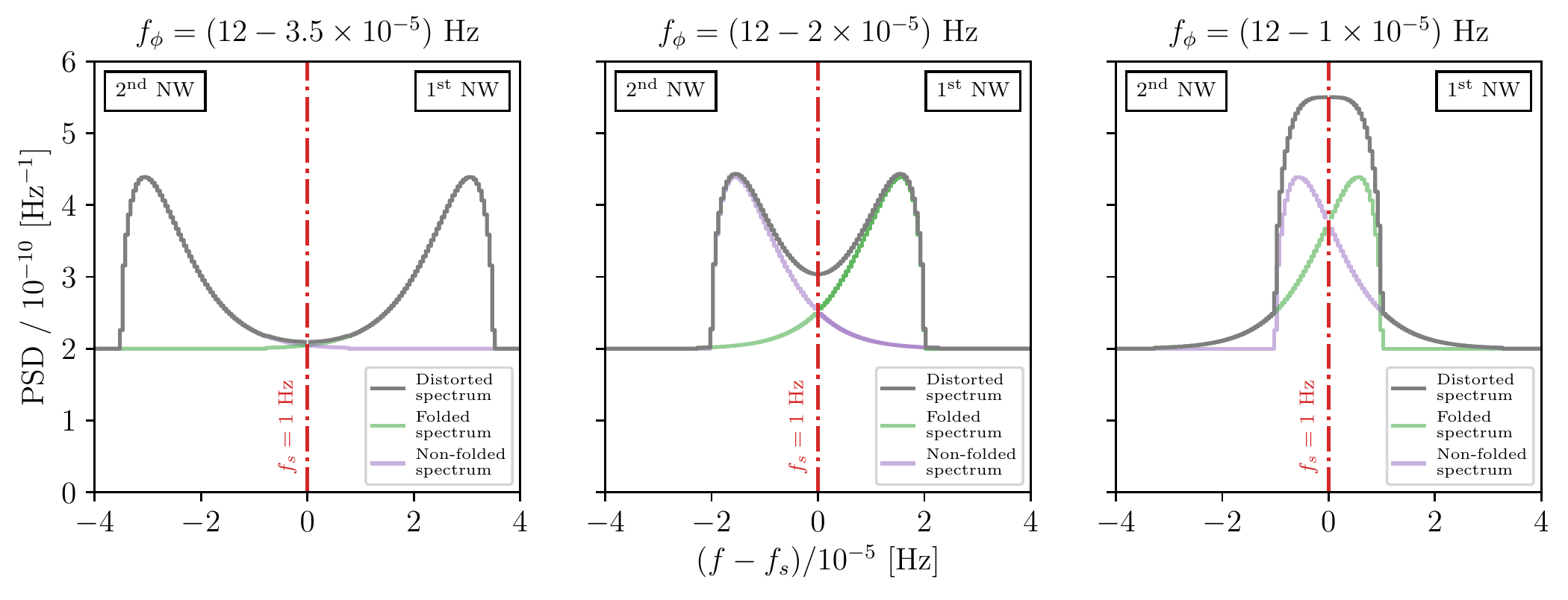}
    \caption{Spectral distortion on ULDM signals close to an integer multiple of the Nyquist frequency, which we set to $0.5$~Hz. We show this effect for three different signals with $d_\phi^2 = 2\times10^{-8}$: $f_\phi = (12-3.5\times 10^{-5})$~Hz (left panel), $f_\phi = (12-2\times 10^{-5})$~Hz (central panel) and $f_\phi = (12-1 \times 10^{-5})$~Hz (right panel). In purple, we show the non-folded spectrum, which is aliased to the second Nyquist window ($2^\mathrm{nd}$~NW) and leaks into the first Nyquist window ($1^\mathrm{st}$~NW); in green we show the folded spectrum, which is aliased to the $1^\mathrm{st}$~NW and leaks into the $2^\mathrm{nd}$~NW; and in grey we show the distorted aliased spectrum, which consists of the sum of the folded and non-folded spectra. We assume the experimental values stated in Table~\ref{table: experimental parameters} and $N_\mathrm{stacks} = 50$. We see that, the closer an aliased signal is imaged to an integer multiple of the Nyquist, the greater the degree of spectral distortion.}
    \label{fig:spec_distortion}
\end{figure}

Spectral distortions owing to folding would be of interest to broadband interferometers for high-frequency ULDM signals well below $f_\phi \sim 10^6$~Hz and sufficiently close to integer multiples of the Nyquist frequency. This is illustrated in Fig.~\ref{fig:spec_distortion} where we show the effect of spectral distortions on three different ULDM signals with $d_\phi^2 = 2\times10^{-8}$ and: $f_\phi = (12-3.5\times 10^{-5})$~Hz, $f_\phi = (12-2\times 10^{-5})$~Hz and $f_\phi = (12-1 \times 10^{-5})$~Hz, which we show in the left, central and right panels respectively. For our choice of sampling frequency, $f_s = 1$~Hz, the non-folded alias of the true ULDM signal is imaged in the second Nyquist window, while the folded alias of the true ULDM signal is imaged in the first Nyquist window. The degree of distortion increases as $f_\phi$ tends to 12~Hz, which corresponds to an even multiple of the Nyquist frequency.  

\begin{figure}
    \centering
    \includegraphics[width = 0.49\textwidth]{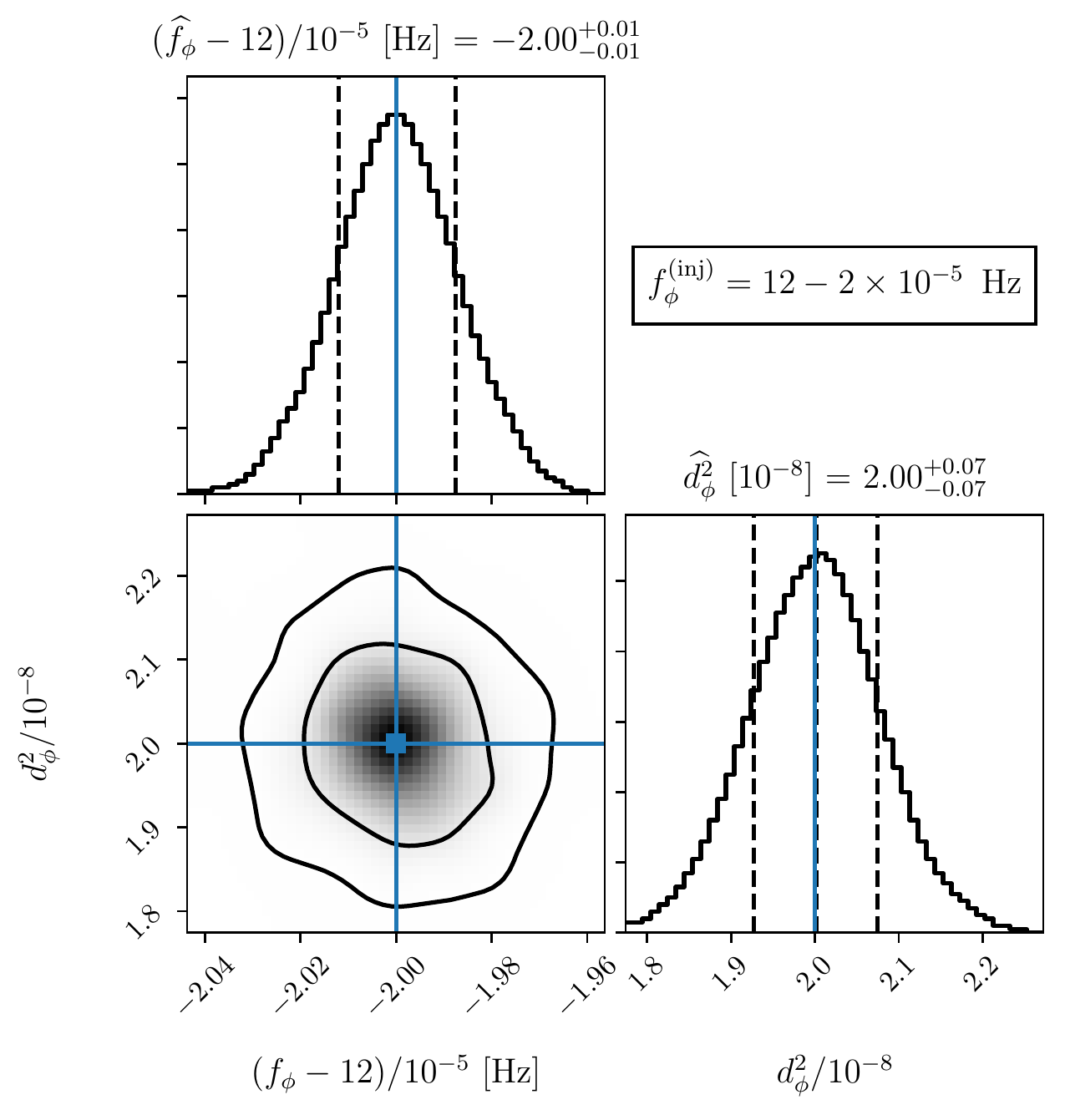}
    \includegraphics[width = 0.49\textwidth]{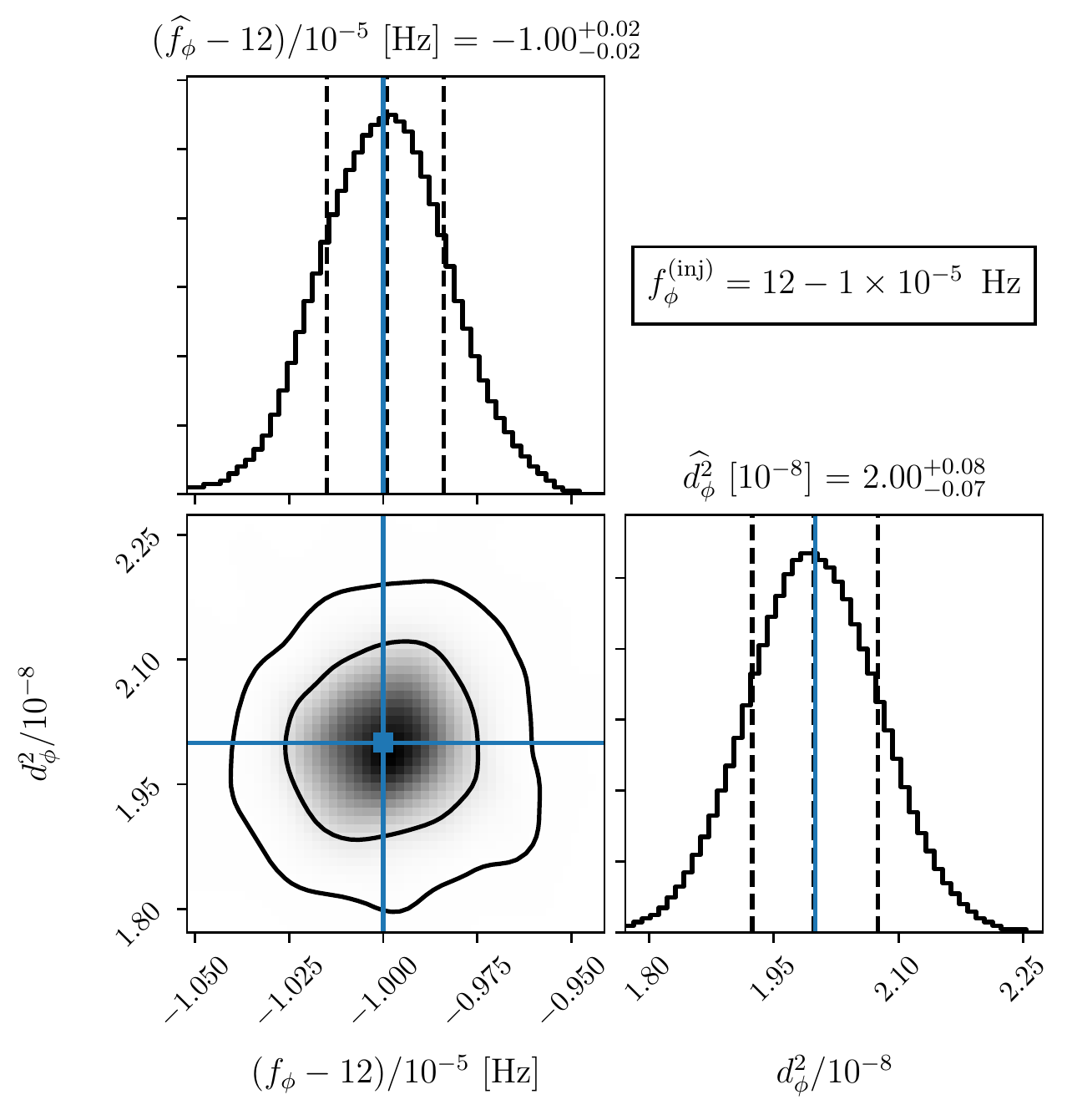}
    \caption{Comparison between the posterior distributions of $f_\phi$ and $d_\phi^2$ for Asimov data sets containing an injected signal with $d_\phi^2 = 2\times10^{-8}$ and $f_\phi = (12-2\times 10^{-5})$~Hz (left panel) and $f_\phi = (12-1 \times 10^{-5})$~Hz (right panel), which correspond to the grey curves plotted in the central and right panels of Fig.~\ref{fig:spec_distortion}, respectively. The injected signal parameters are shown by blue solid lines in 1D and a blue square in 2D planes. The $1\sigma$ and $2\sigma$ credible regions are shown with solid lines. The inferred ULDM parameters are consistent with the injected signal to within $2\sigma$ credible region independently of the degree of spectral distortions due to folding.}
    \label{fig:reconstructed parameters distortion}
\end{figure}

By taking into account spectral distortions, the likelihoods defined in section~\ref{subsec: likelihood} can correctly reconstruct such signals, independently of the degree of distortion induced by folding. We show this in Fig.~\ref{fig:reconstructed parameters distortion}, where we plot the posterior distribution of the reconstructed ULDM parameters, namely $f_\phi$ and $d_\phi^2$, based on Asimov data sets containing signals with various degrees of distortion characterised by: $f_\phi = (12-2\times10^{-5})$~Hz and $d_\phi^2 = 2\times10^{-8}$, and $f_\phi = (12-1\times10^{-5})$~Hz and $d_\phi^2 = 2\times10^{-8}$, which correspond to the distorted signals shown in the central and right panels of Fig.~\ref{fig:spec_distortion}, respectively. In each case, the injected signal parameters are correctly reconstructed: for the injected signal at $f_\phi = (12-2\times10^{-5})$~Hz, for which we show the posterior in the left panel of Fig.~\ref{fig:reconstructed parameters distortion}, we infer $\widehat{f_\phi} = (12-2^{+0.01}_{-0.01}\times10^{-5})$~Hz and $\widehat{d_\phi^2} = 2^{+0.07}_{-0.07} \times 10^{-8}$; for the injected signal at $f_\phi = (12-1\times10^{-5})$~Hz, for which we show the posterior in the right panel of Fig.~\ref{fig:reconstructed parameters distortion}, we infer $\widehat{f_\phi} = (12-1^{+0.02}_{-0.02}\times10^{-5})$~Hz and $\widehat{d_\phi^2} = 2^{+0.08}_{-0.07} \times 10^{-8}$. 

Interestingly, while the uncertainty on $\widehat{d_\phi^2}$ is the same for both signals, the uncertainty on the mass differs appreciably: for the injected signal at $f_\phi = (12-2\times10^{-5})$~Hz, the relative uncertainty on $(\widehat{f_\phi}-12)$~Hz is $0.5\%$, while for the injected signal at $f_\phi = (12-1\times10^{-5})$~Hz, the relative uncertainty on $(\widehat{f_\phi}-12)$~Hz is $\approx 2\%$. This difference can be explained by noting that the degree of distortion in the signal's line-shape is highly sensitive to the value of the alias of $f_\phi$ with respect to the characteristic line-width $\Delta f_\phi$. For signals satisfying $  \Delta f_\phi \ll f_s - f_\phi^* < f_\phi \, v_\mathrm{esc}^2/2 $ (e.g. the $f_\phi = (12-2\times10^{-5})$~Hz signal shown in the central panel of Fig.~\ref{fig:spec_distortion}), the distortion will affect the high-tail of the imprinted dark matter speed distribution. Since this deviation from the SHM cannot be accounted for by tuning~$d_\phi^2$, such signals would be more sensitive to changes in $f_\phi^*$, and thus $f_\phi$.

In summary, we conclude that:~\textit{i}) super-Nyquist signals that are affected by spectral distortions due to folding can be correctly reconstructed using the tools presented in section~\ref{sec: freq-dom analysis}; and \textit{ii}) the relative uncertainty on the reconstructed coupling $d_\phi^2$ is independent of the degree of distortion, while the relative uncertainty on the reconstructed value of $f_\phi$ is smallest for signals satisfying $  \Delta f_\phi \ll f_s - f_\phi^* < f_\phi \, v_\mathrm{esc}^2/2 $, i.e., distorted signals whose deviation from their corresponding non-distorted spectra predominantly affects the high-speed tail of the imprinted dark matter speed distribution.

\subsection{Example discovery search}\label{subsec: example search}

We complete this section with an example of a discovery analysis, which will provide a unified context for the phenomena related to aliasing discussed earlier.

For the purpose of comparison, we generate two MC data sets\footnote{Here, we depart from the Asimov data set, of which we made ample use in previous sections. Additionally, for the purposes of comparison, we choose the same seed for both data sets, so that fluctuations about the expected signal are the same in both cases.} in the frequency domain (see Appendix~\ref{app:MC_sims} for details) assuming a ULDM signal with $f_\phi = 9.1$~Hz, $d_\phi^2 = 10^{-9}$, the experimental parameters mentioned in Table~\ref{table: experimental parameters} and two different sampling frequencies: 0.3~Hz and 3~Hz. To maximise the degree to which folded and non-folded signals can be disentangled, while also minimising the number of bins over which to perform the scan (i.e.~maximising $N_\mathrm{stacks}$ at fixed $T_\mathrm{int}$), we set $N_\mathrm{stacks} = 100$ (cf.~section~\ref{subsec: folding}). 

To reconstruct the injected signal parameters, we perform a parameter scan in the $(m_\phi,\,d_\phi^2)$ plane of interest 
using a nested sampling algorithm as implemented in \textsf{PyMultiNest} \cite{Buchner:2014nha}, a Python interface to \textsf{MultiNest} \cite{Feroz:2007kg,Feroz:2008xx,Feroz:2013hea}. The range of ULDM masses (or frequencies) that are scanned over also includes multiples of the first and second Nyquist windows. In this example, we restrict the analysis to frequencies between $\sim 2$~Hz and $\sim 22$~Hz, for which we expect the reconstructed signals to not be contained within the C.R.~at the 95.4\% C.L. of the best-fit point for a $5.1\sigma$ locally ($0.4\sigma$ globally) significant discovery (i.e.~$\mathrm{TS}_\mathrm{max} \approx 26$, cf.~section~\ref{subsec: distinguishing aliased sigs}). Since $N_\mathrm{stacks} > 1$, the test statistic for discovery is evaluated for the stacked likelihood defined in Eq.~\eqref{eqn:stacks_like}. Additionally, to disentangle aliased from non-aliased signals, the ULDM model used in the scan is defined for fixed ULDM speed parameters, specifically $v_0$, $v_\mathrm{obs}$ and $v_\mathrm{esc}$, which we set to those of the SHM (cf.~section~\ref{subsec: alias vs nonalias}).

\begin{figure}[t]
    \centering
     \includegraphics[width=0.9\textwidth]{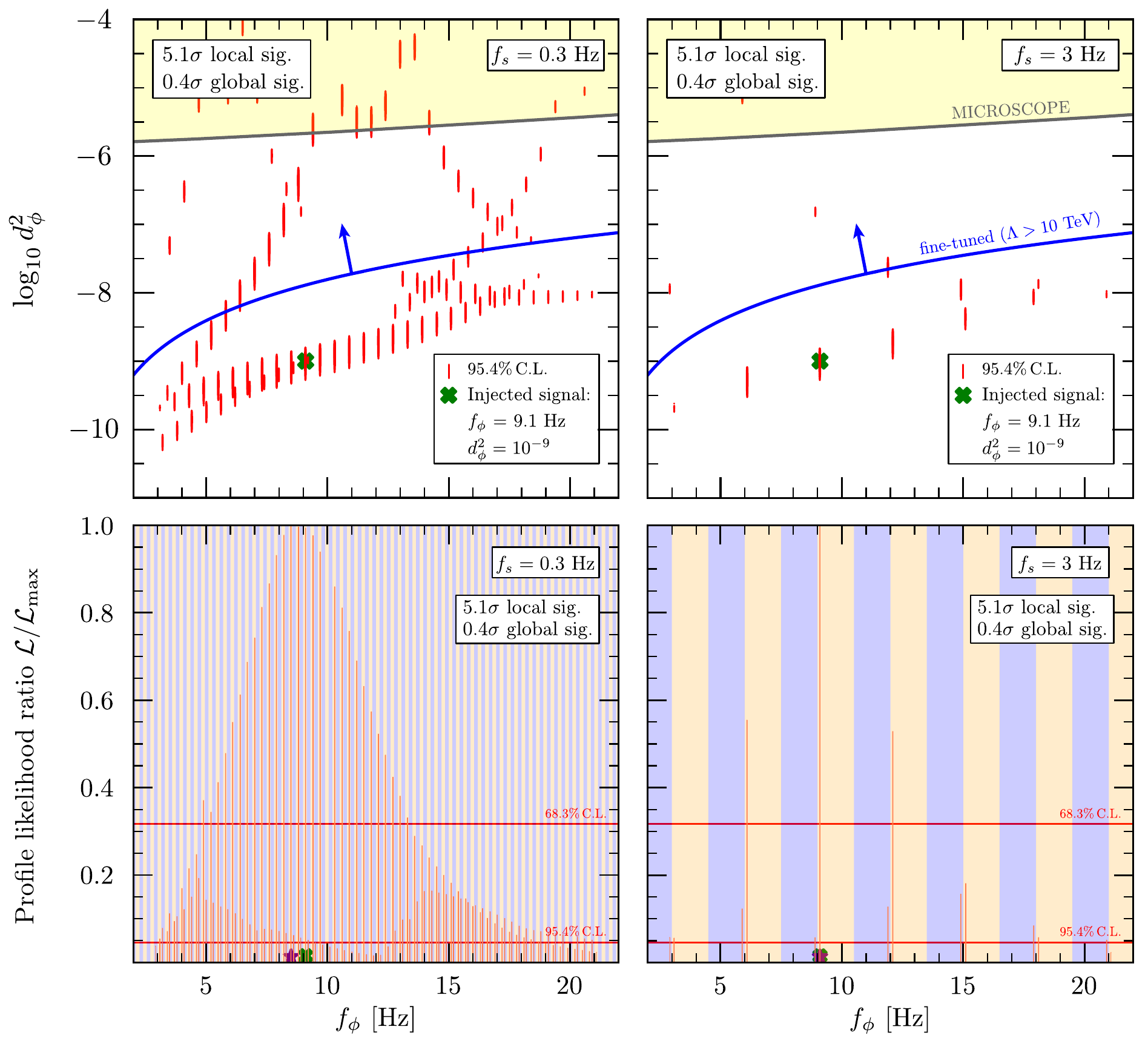}
    \caption{Discovery analysis on a MC-generated data set containing an injected signal with $f_\phi = 9.1$~Hz and $d_\phi^2 = 10^{-9}$ for $f_s = 0.3$~Hz (left column) and $f_s = 3$~Hz (right column). In the upper row, we show the islands of parameter space that are consistent with the inferred best-fit at the 95.4\% C.L. in red, and the injected signal with a green cross. The yellow shaded region has been excluded by MICROSCOPE, while the fine-tuned region of parameter space assuming a UV-cutoff $\Lambda = 10$~TeV is located above the blue line. The lower row shows the one-dimensional profile likelihood ratio for the ULDM frequency. The beige (blue) shaded regions mark multiples of the first (second) Nyquist window. The horizontal magenta lines correspond to the C.R. at the $68.3\%$ and $95.4\%$ C.L. In these panels, we label the best-fit point with a purple star.} 
    \label{fig: 1D and 2D PLRs}
\end{figure}

In Fig.~\ref{fig: 1D and 2D PLRs}, we plot the results of this analysis for both data sets. In the upper row, we display the regions of parameter space that are consistent with the global maximum at the 95.4\% C.L. for $f_s = 0.3$~Hz and $f_s = 3$~Hz, which we show on the left and right, respectively. For $f_s = 0.3$~Hz, we see that most Nyquist windows between $\sim 2$~Hz and $22$~Hz are consistent with the best-fit value, which is identified at $\approx 8.5$~Hz and is consistent with the injected signal. For $f_s = 3$~Hz, however, fewer regions of parameter space are consistent with the best-fit, which is correctly reconstructed at $\approx 9.1$~Hz. Different from other dark matter direct detection experiments, these regions of parameter space are disconnected, precisely in light of the symmetry between Nyquist windows. For $f_s = 3$~Hz, however, fewer regions of parameter space are consistent with the best-fit value, as a result of aliasing: because the peak in the first Nyquist window is located at $\approx 0.1$~Hz, scanned frequencies $f_\phi^\mathrm{scan}$ satisfying $0.5 \lesssim f_\phi^\mathrm{scan}/f_\phi \lesssim 2$ and identified at 0.1~Hz are consistent with the injected signal; in agreement with the conclusions of section~\ref{subsec: distinguishing aliased sigs}, since $f_\phi^\mathrm{scan} = f_\phi + a f_\mathrm{Ny}$, for integers $a$, the smaller $f_\mathrm{Ny}$, and so $f_s$, the more disconnected regions of parameter space consistent with the results.

In the lower row of Fig.~\ref{fig: 1D and 2D PLRs}, we illustrate the profile likelihood ratio with respect to $f_\phi$ for the corresponding data sets, which is defined in terms of the discovery test statistic as 
\begin{equation}
\mathcal{L}(f_\phi)/\mathcal{L}_\mathrm{max} = \exp \left \{ \left [\mathrm{TS}\left (f_\phi, \widehat{d_\phi^2}\right)-\mathrm{TS}\left(\widehat{f_\phi, d_\phi^2}\right) \right ]/2 \right\} \, . 
\end{equation}
Here, we shade the range of frequencies that would be aliased to the first Nyquist window in beige, and the range of frequencies that would be aliased to the second Nyquist window in blue. In these panels we see that a peak is visible in each scanned Nyquist frequency range, which implies that a signal is reconstructed in all scanned multiples of the Nyquist frequency. However, not all of these signals will be compatible with the global maximum at $2\sigma$. Indeed, disconnected islands are only visible in the corresponding mass-coupling plane when the profile likelihood ratio exceeds the $95.4\%$ C.R.

\begin{figure}[t]
    \centering
     \includegraphics[width=0.9\textwidth]{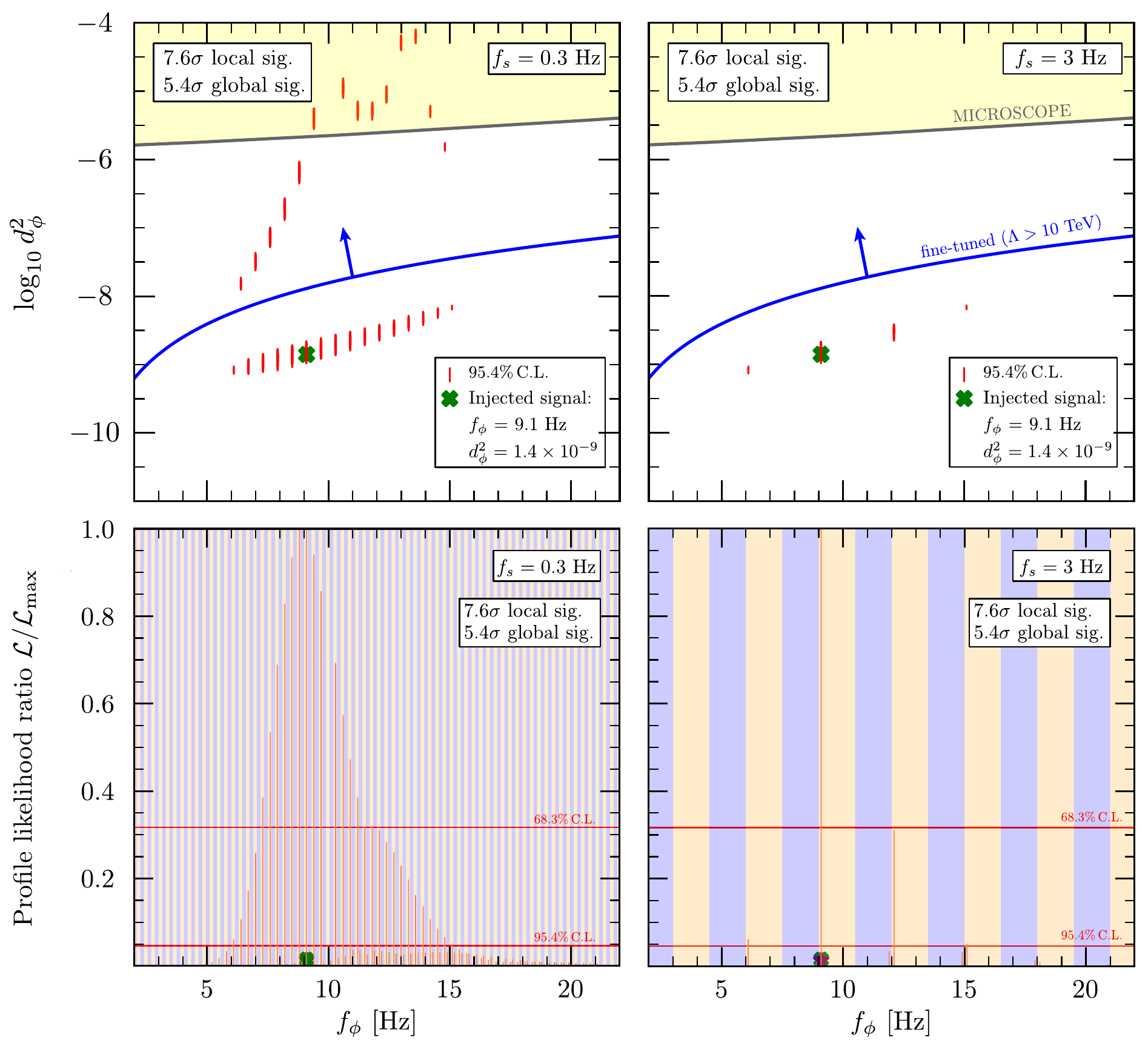}
    \caption{Discovery analysis on a MC-generated data set containing an injected signal with $d_\phi^2 = 1.4\times 10^{-9}$ and other parameters as in Fig.~\ref{fig: 1D and 2D PLRs}. The upper row shows (in red) the islands of parameter space that are consistent with the inferred best-fit value at the 95.4\% C.L. The lower row shows the one-dimensional profile likelihood ratio for the ULDM frequency, where the beige (blue) shaded regions mark multiples of the first (second) Nyquist window. The horizontal magenta lines correspond to $68.3\%\,(1\sigma)$ and $95.4\%\,(2\sigma)$ confidence region of the right panels. The injected signal is shown with a green cross while the best-fit point is labelled with a purple star.}
    \label{fig: 1D and 2D PLRs 2}
\end{figure}

To illustrate the impact of small changes in the size of $d_\phi^2$ on the ability to correctly reconstruct super-Nyquist ULDM signals, in Fig.~\ref{fig: 1D and 2D PLRs 2}, we plot the results of a discovery analysis on two MC data sets which contain a ULDM signal with $f_\phi = 9.1$~Hz and $d_\phi^2 = 1.4\times10^{-9}$, the latter differing from the value of $d_\phi^2$ used in generating the data sets for Fig.~\ref{fig: 1D and 2D PLRs} by $\sqrt{2}$.
All other parameters are identical to the ones that were used to generate the results illustrated in Fig.~\ref{fig: 1D and 2D PLRs}. In agreement with the results from sections~\ref{subsec: folding} and~\ref{subsec: distinguishing aliased sigs}, for this choice of coupling a reduced number of Nyquist windows will contain signals that are consistent with the best-fit value, which is identified at $\approx 9.1$~Hz for both $f_s = 0.3$~Hz and $f_s = 3$~Hz and consists of a $7.6\sigma$ locally ($5.4\sigma$ globally) significant discovery. 
Additionally, as shown in the lower panels of  Fig.~\ref{fig: 1D and 2D PLRs 2}, for such a choice of coupling, no range of frequencies whose folded alias would lie in the second Nyquist window is contained within the C.R. at the 95.4\% C.L., independently of the sampling frequency. 
This confirms that folded and non-folded aliased signals can be disentangled when the data contains a highly significant signal (c.f. section~\ref{subsec: folding}). 

In light of the ULDM signal's dependence on experimentally tuneable parameters, it also follows that the discovery analysis of super-Nyquist signals is highly sensitive to small changes in sequence parameters. For example, the scalar ULDM signal amplitude in gradiometer experiments depends linearly on the gradiometer length $\Delta z$ and the number of LMT pulses $n$. 
Therefore, if a super-Nyquist signal is detected with a $5\sigma$ local significance, it is possible to improve the detection significance to $5\sigma$ global significance, without changing the frequency associated with the experiment's peak sensitivity, by increasing $\Delta z$ by a factor of $2^{1/4}$.\footnote{In the regime $T_\mathrm{int}^\mathrm{st} >\tau_c$ considered here, the scalar ULDM signal amplitude also scales with the number of LMT pulses and interrogation time as $n$ and $T^{5/4}$, respectively~\cite{Badurina:2021lwr}. Changing these parameters, however, changes the frequency at which the experiment's sensitivity peaks. In the super-Nyquist regime considered here, any small change in these parameters may dramatically reduce the size of the signal amplitude around the best fit point. Hence, changing these parameters \textit{post}-discovery of a $5\sigma$ signal should be done with care.} We therefore recommend that future broadband atom gradiometer experiments consider implementing designs that could be modified post-construction. 

\begin{figure}[t!]
    \centering
    \includegraphics[width=0.9\textwidth]{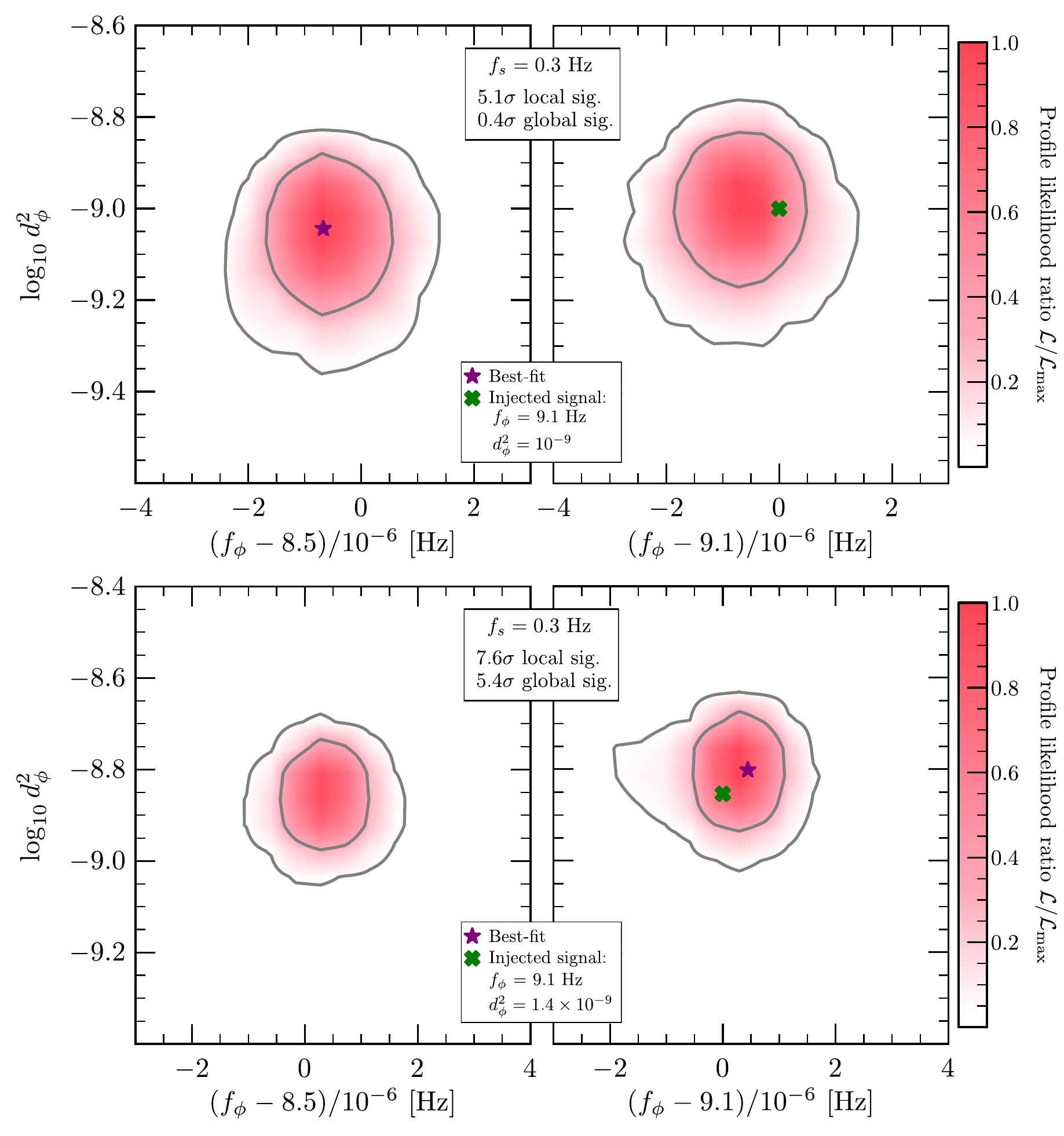}
    \caption{Enlargement of the parameter space shown in Figs.~\ref{fig: 1D and 2D PLRs} and \ref{fig: 1D and 2D PLRs 2} close to 8.5~Hz (left column) and 9.1~Hz (right column) for $f_s = 0.3$~Hz when the injected signal is at $d_\phi^2 = 10^{-9}$ (top row) and $d_\phi^2 = 1.4 \times 10^{-9}$ (bottom row). In both cases, the mass resolution is on the order of the stacked frequency resolution. The contours, which correspond to the confidence regions at the 68.3\% and 95.4\% C.L., are tightest for the most significant injected signal.}
    \label{fig:selected windows}
\end{figure}

Finally, for completeness, in Fig.~\ref{fig:selected windows} we enlarge the C.R. for the $f_s = 0.3$~Hz case in the vicinity of 8.5~Hz and 9.1~Hz, for both choices of coupling. For the data set containing a signal with $d_\phi^2 = 10^{-9}$, we observe that, despite being four Nyquist windows away from the injected mass, the reconstructed ULDM mass is still consistent with 9.1~Hz at the $68.3$ C.L. This is to be contrasted with the dataset containing a signal with  $d_\phi^2 = 1.4\times10^{-9}$, for which the best fit point lies in the vicinity of 9.1~Hz. Additionally, in both cases the resolution on the mass is on the order of the stacked frequency resolution $\Delta f^\mathrm{st} = 10^{-6}$~Hz. 
In agreement with the notion of the $\mathrm{TS}$ as a measure of significance, the contours of the more significant signal (lower panels) are tighter.

\section{Discussion and Summary}\label{sec: summary}

Super-Nyquist ULDM signals, which we defined as ULDM signals whose spectral content exceeds half of an experiment’s sampling frequency, are a well-motivated target for future broadband atom gradiometer experiments. These signals, however, would be affected by spectral features that deviate substantially from those of sub-Nyquist signals, and so would not be correctly reconstructed using the analysis routines previously discussed in the literature. To address this lacuna, in this work we have provided the first systematic approach to discovering super-Nyquist ULDM signals with broadband atom gradiometers through the use of a comprehensive likelihood formalism and statistical framework. 

To this end, we have conducted a detailed exploration of the phenomenon of aliasing, whereby any super-Nyquist signal is identified at a frequency between zero and the Nyquist frequency.  Importantly, even after aliasing to a lower frequency, the width of the spectral lineshape, which exhibits a linear dependence on the ULDM mass (or, equivalently, the ULDM frequency), remains unchanged. As explained in section~\ref{subsec: alias vs nonalias}, this characteristic enables the differentiation of aliased signals from non-aliased ones, provided that the experimental frequency resolution, which is determined by the integration time and the number of stacks (cf.\ table~\ref{tab:Dictionary}), is sufficiently high.

Furthermore, in addition to the shift to lower frequencies, we discussed two other aspects of aliasing: folding and distortion. 
Folding refers to the reflection of the signal around the Nyquist frequency, while distortion occurs when the spectral content of a signal exceeds the frequency range of a given Nyquist window and is added onto the original signal spectrum, resulting in a substantially modified lineshape (cf.\ sections~\ref{subsec: folding}-\ref{subsec: spectral distortions}). 
By accounting for all these aspects of aliasing, we have shown that our likelihood formalism can give an accurate reconstruction of the original signal parameters, as long as the frequency resolution is large enough. Indeed, 
we found that an experimental (stacked) frequency resolution greater than approximately five times the signal linewidth is sufficient.

A notable feature that occurs in the reconstruction of super-Nyquist signals was demonstrated in Figs.~\ref{fig: 1D and 2D PLRs} and~\ref{fig: 1D and 2D PLRs 2}.
Because ULDM frequencies that differ by integer multiples of the sampling frequency are identified at the same aliased frequency, the discovery analysis recovers discrete islands of parameter space. 
Each island represents a set of ULDM frequencies consistent with the best fit point.
Within each island, ULDM frequencies of order the experimental (stacked) frequency resolution are found to be consistent with the signal (cf.\ Fig.~\ref{fig:selected windows}),
while the overall number of islands depends on the statistical significance of the ULDM signal, in conjunction with the magnitude of the sampling frequency.

Our systematic exploration of the phenomenon of aliasing has shown that the ability to accurately reconstruct super-Nyquist ULDM signals depends primarily on experimentally tunable parameters that can be set pre- or post-data taking. These include, respectively, the sampling frequency and the sequence parameters, which affect the significance of the ULDM signal for a given ULDM coupling value, and the experimental stacked frequency resolution. These considerations may inform future experimental designs and enhance ULDM detection strategies with upcoming atom gradiometers.

There is scope to go beyond the analysis presented in this work in two primary ways.
Firstly, we have neglected sources of coloured noise, such as gravity gradient noise (GGN), which could dominate the background at both frequencies below and above the Nyquist frequency.
For example, at frequencies below $\sim 0.5$~Hz, GGN is expected to eventually dominate the background of terrestrial long-baseline experiments~\cite{MIGAconsortium:2019efk, Badurina:2022ngn}.
Therefore, high-frequency signals of a ULDM nature that are aliased to low frequencies, especially below 0.5~Hz, would have to contend with a frequency-dependent background of a geological nature. 
To accurately model this background, anti-aliasing filtering techniques like those proposed in Ref.~\cite{PhysRevE.71.066110} could be implemented to distinguish between the aliased and non-aliased parts of the background spectrum. 
By leveraging the dependence of the GGN contribution on the ground's vertical spectrum at the Earth's surface and on local geological properties, 
the filter could be modelled on local density measurements and seismometer data~\cite{Arduini:2023wce}. 

Furthermore, coloured noise that dominates the background above the Nyquist frequency would suffer from aliasing and related effects discussed in this work. 
Coloured noise whose spectrum leaks beyond the Nyquist frequency will be folded to lower frequencies and added to the non-aliased low-frequency spectrum. 
Therefore, to accurately reconstruct a super-Nyquist ULDM signal in the presence of this background, 
it would be necessary to subtract the expected non-aliased low-frequency background spectrum from the background model.
Alternatively, to avoid introducing systematic errors, it may be feasible to design gradiometer sequences that are simultaneously sensitive to the non-aliased low-frequency spectrum signal and insensitive to the ULDM signal. In this case, a noise-free spectrum could be obtained through spectral subtraction~\cite{vaseghi2008advanced}. A more detailed investigation in this direction is left for future work.

In this work, we have also assumed that the sampling frequency of the experiment is constant and known with arbitrary precision. 
This assumption, which presents an idealised scenario, leads to the second primary extension of this work. 
Interestingly, as proven in Ref.~\cite{refId0}, in the case of unevenly sampled, the Nyquist frequency is given by $1/(2\Omega)$, where $\Omega$ is the largest factor such that the temporal spacing between any sampled point is given by an integer multiple of $\Omega$. 
Hence, choosing $\Omega = 0.1$~s and performing $T_\mathrm{int}/\Omega$ measurements at times $t_i = n_i \Omega $, where $n_i$ is a uniformly sampled integer between zero and $T_\mathrm{int}/\Omega$, the Nyquist frequency would be given by $f_\mathrm{Ny} = 10$~Hz. 
Following this argument, measurements that are performed at different times with finite precision would be characterised by a very large Nyquist frequency~\cite{10.1111/j.1365-2966.2006.10762.x}. For example, assuming a timing precision on the order of $10^{-3}$~s, the Nyquist frequency will be bounded above by $5\times 10^{2}$~Hz, where this bound will be reached in the limit that no larger factors exist. 
Since the measurement of the atom populations at the end of the experiment is dictated in large part by the timing of lasers, we expect this timing precision to be achievable. 
While this technique holds promise for completely eliminating aliasing effects from atom gradiometer data, its implementation would require the application of the Lomb-Scargle periodogram~\cite{VanderPlas_2018}, 
which would modify both the statistical features of the signal as well as the analytical form of the signal and background. Therefore, further investigation of this approach is deferred to future studies.

Finally, we emphasise that while our discussion has been specifically tailored to scalar ULDM searches, the analysis and findings of this work can be readily adapted  to other ULDM searches utilising atom gradiometers, such as spin-1 DM~\cite{Graham:2015ifn}, provided that the assumptions made here remain valid. 
More generally, the conclusions of this work would also be relevant to other state-of-the-art broadband experiments hunting for ULDM candidates, including broadband sensors searching for axion-like particles.

\acknowledgments
We are grateful to members of the AION Collaboration for many fruitful discussions and to John Carlton and John Ellis for comments on the manuscript. A.B. thanks Tom\'as Gonzalo, Anders Kvellestad and Andre Scaffidi for helpful discussions. L.B.\ acknowledges support from the Science and Technology Facilities Council (STFC) Grant No. ST/T506199/1. A.B.\ and C.M.\  are supported from the STFC Grant No.~ST/T00679X/1. This work made use of \textsf{GNU Parallel}~\cite{tange_2023} and the \textsf{Matplotlib} \cite{Hunter:2007}, \textsf{SciPy} \cite{2020SciPy-NMeth} and \textsf{Numpy} \cite{harris2020array} modules of the \textsf{Python} \cite{10.5555/1593511} package. Plots were generated using the \textsf{Corner}~\cite{corner} and \textsf{pippi~v2.2}~\cite{Scott:2012qh} packages. For the purpose of open access, the authors have applied a Creative Commons Attribution (CC BY) license to any Author Accepted Manuscript version arising from this submission. No experimental data sets were generated by this research.

\appendix


\section{Deriving aliasing from the properties of the Fourier transform}\label{app: symmetries of the FT}

In this appendix, we will show how aliasing arises from the properties of the Fourier transform of a continuous time-dependent signal, which is sampled in the time-domain. For the sake of clarity and generality, we will present an argument that is independent of the exact form of the ULDM signal discussed in this work. 

Let $\Phi(t)$ be a continuous time-dependent signal that we wish to measure and analyse in the frequency domain. Let us also assume that the signal is sampled at a rate $f_s = 1/\Delta t$ for a time $T_\mathrm{int} \rightarrow \infty$, i.e., our time series is infinitely long.\footnote{We take this limit to avoid spectral leakage due to windowing and to simplify the derivation.}~The sampled time-dependent signal can then be written as $\Phi_s(t) = \Phi(t) \Sh(t)$, where $\Sh(t)$ is the Dirac comb and is defined as
\begin{equation}
\Sh(t) = \sum_{m = -\infty}^{\infty} \delta (f_s t - m)\,.
\end{equation}
Hence, the Fourier transform of the sampled signal $\Phi_s(t)$ takes the form
\begin{equation}\label{app:eq FT of sampled signal}
\widetilde{\Phi_s}(f) =  \int_{-\infty}^{\infty} \Phi_s(t) \, e^{-2\pi i f t} \, dt = \int_{-\infty}^{\infty} \Phi(t) \, \Sh(t) \, e^{-2\pi i f t} \, dt \,.
\end{equation}
After making use of the Fourier series expansion of the Dirac comb, namely $\Sh(t) = \sum_{\kappa \, = \,-\infty}^{\infty}e^{2\pi i \kappa f_s t}$~\cite{EPFL}, Eq.~\eqref{app:eq FT of sampled signal} can be written as
\begin{equation}
\widetilde{\Phi_s}(f) = \sum_{\kappa \, = \,-\infty}^{\infty} \int_{-\infty}^{\infty} \Phi(t) \, e^{-2\pi i (f + \kappa f_s)t} \, dt = \sum_{\kappa \, = \, -\infty}^{\infty}  \widetilde{\Phi}(f+\kappa f_s) \,.
\end{equation}
That is, the Fourier transform of the sampled signal at $f$ corresponds to the sum of the Fourier transform of $\Phi(t)$ at $f$ and all of its aliases. Using the notation of Ref.~\cite{PhysRevE.71.066110}, we have
\begin{equation}
\widetilde{\Phi_s}(f) = \widetilde{\Phi}(f) + \sum_{\kappa \, \neq \, 0} \widetilde{\Phi}(f+\kappa f_s) \,.
\end{equation}
In turn, this implies that the power spectral density (PSD) of the sampled signal $\Phi_s$ is (up to a normalisation factor)
\begin{align}\label{app:eq: aliased PSD}
S_s(f) &=  \left |\widetilde{\Phi_s}(f) \right |^2 \nonumber \\
&= \left |\widetilde{\Phi}(f) \right |^2  + \sum_{\kappa \, \neq \, 0} \left | \widetilde{\Phi}(f + \kappa f_s) \right |^2 \nonumber \\
&\hspace{5mm} + 2 \sum_{\kappa \, \neq \, 0} \mathrm{Re} \left \{\widetilde{\Phi}(f) \widetilde{\Phi}^*(f+\kappa f_s) \right \} +2 \sum_{\substack{\kappa \, \neq \, \kappa' \\ \kappa,  \,\kappa' \, \neq \, 0}} \mathrm{Re} \left \{\widetilde{\Phi}(f+\kappa f_s) \widetilde{\Phi}^*(f+\kappa' f_s) \right \} \, .
\end{align}

In the case of stochastic signals, it is natural to consider the expectation value of the PSD. In particular, let us assume that the signal of interest is a superposition of Fourier modes, each of which is characterised by a random phase that is uniformly sampled between 0 and $2\pi$. In this case, as the phases of different Fourier modes are uncorrelated and independent, the expectation value of the terms in the last line of Eq.~\eqref{app:eq: aliased PSD} vanish, such that
\begin{equation}\label{app:eq: PSD alias}
\begin{aligned}
\left \langle S_s(f) \right \rangle &= \left \langle \left |\widetilde{\Phi}(f) \right |^2 \right  \rangle +  \sum_{\kappa \, \neq \, 0}  \left \langle \left |\widetilde{\Phi}(f+\kappa f_s) \right |^2 \right  \rangle \\
&= \left \langle S(f) \right \rangle + \sum_{\kappa \, \neq \, 0} \left \langle S(f+\kappa f_s) \right \rangle \,.
\end{aligned}
\end{equation}
That is, the expected value of the PSD of the sampled signal at $f$ is the PSD of the continuous function $S(f)$ plus the sum of all of its aliases. As $\widetilde{\Phi}(f) = \widetilde{\Phi}^*(-f)$, $\widetilde{\Phi}(f - |\kappa| f_s) = \widetilde{\Phi}^*(-f + |\kappa| f_s)$. Thus, we can rewrite Eq.~\eqref{app:eq: PSD alias} as a sum over positive frequencies (i.e., $f > 0$ and $\kappa \geq 1$), i.e.
\begin{equation}
\left \langle S_s(f) \right \rangle = \left \langle S(f) \right \rangle + \sum_{\kappa \, \geq \, 1} \left \langle S(f+\kappa f_s) \right \rangle + \left \langle S(-f + \kappa f_s) \right \rangle \,.
\end{equation}
This completes the derivation. Importantly, we made no assumption concerning the form of the amplitude of the signal, so the results shown here apply for all stochastic signals affecting an experiment. Within the context of atom gradiometer experiments, these include the scalar ULDM signal and coloured noise, such as gravity-gradient noise (GGN).

White noise, e.g., atom shot noise, however, cannot be described using the above analysis. Indeed, as white noise, by definition, is non-zero at all frequencies, its aliased spectrum would be infinite. Therefore, when considering shot-noise limited experiments, we will assume that the noise spectrum is band-limited, which implies that aliasing and folding do not affect the background.

\section{Test statistic for setting upper limits}\label{app: q for upper lims}

In this appendix, we study the test statistic for setting upper limits on the ULDM-SM coupling at particular mass values. 
As before, we set $\boldsymbol{\theta}_{\mathrm{sig}} = (m_{\phi},\,d^2_\phi)$, where we remind the reader that $m_{\phi}$ is the ULDM mass and $d_\phi$ is the coupling strength of the relevant linear interaction between ULDM and SM operators.~Thus, we can define the test statistic for setting upper limits on the ULDM-SM coupling $d_\phi$ as
\begin{equation}\label{eqn:TS_upper_lims}
    q\left(m_{\phi},\,d^2_\phi\right) = 
    \begin{cases} 
        2 \ln{\dfrac{\mathcal{L} \big(d \mid \mathcal{M}_\mathrm{S\,+\,B}, \big\{\widehat{\widehat{\boldsymbol{\theta}}}_\mathrm{nuis}, \, m_\phi,\,d^2_\phi \big\} \big)}{\mathcal{L} \big(d \mid \mathcal{M}_\mathrm{S\,+\,B} \,, \big\{  \widehat{\boldsymbol{\theta}}_\mathrm{nuis}, \, m_\phi,\,\widehat{d^2_\phi} \big\} \big)}} & d^2_\phi \geq \widehat{d^2_\phi} \,, \\[1.5mm]
        0 & d^2_\phi < \widehat{d^2_\phi} \,,
    \end{cases}
\end{equation}
where $\widehat{d^2_\phi}$ is the value of $d^2_\phi$ that maximises the likelihood at fixed $m_\phi$. Here, $\widehat{\boldsymbol{\theta}}_{\text {nuis}}$ denotes the values of the nuisance parameters that maximise the likelihood in the signal-plus-background hypothesis given the best-fit value of the squared coupling strength $\widehat{d^2_\phi}$ (i.e., the denominator term); $\widehat{\widehat{\boldsymbol{\theta}}}_{\text {nuis}}$ represents the values of the nuisance parameters that maximise the likelihood in the signal-plus-background hypothesis given the squared coupling strength $d^2_\phi$ (i.e., the numerator term).

In the limit of a large data sample, i.e., $T_\mathrm{int} \gg \tau_c$, the signal is spread over multiple frequency bins. Hence, we can invoke the Wald approximation and Wilks' theorem~\cite{Cowan:2010js}, which imply that the test statistic $q$ at fixed $m_\phi$ is described by a half chi-squared distribution with one degree of freedom~\cite{10.1214/aoms/1177728725}.\footnote{This follows from the fact that we do not consider downward fluctuations of the background as evidence against the background-only hypothesis, i.e., we are effectively testing for $\widehat{d_\phi^2} \geq 0$.}~For a given $m_\phi$, the 95\% confidence level limit on $d^2_\phi$ is set when $q\left(m_\phi, d_{\phi,\,95\%}^{2}\right) \approx -2.706$. In this limit, the $N\sigma$ confidence intervals on $d^2_{\phi,\,95\%}$ can then be computed via
\begin{equation}\label{eqn:N-sigma}
    q \left (m_{\phi},\,d^2_{\phi,\,95\% \,\pm \,N\sigma}\right) = - \left(\Phi^{-1}(0.95) \pm N \right)^2 \,,
\end{equation}
where $\Phi^{-1}$ is the inverse of the cumulative distribution function for the
Normal distribution. For our choice of integration time $T_\mathrm{int} = 10^8$\,s and no stacking, this regime applies for ULDM masses above $\sim 10^{-17}$\,eV, i.e., frequencies above $10^{-2}$~Hz. With mild stacking (e.g. $N_\mathrm{stacks} = 10$), which reduces the value of the integration time by the number of stacks $N_\mathrm{stacks}$, this regime will still apply for signals oscillating at frequencies greater than $\mathcal{O}(0.1\,\mathrm{Hz})$.

\section{Monte Carlo simulations}\label{app:MC_sims}
In this section, we discuss in detail our approach towards generating Monte Carlo (MC) simulations of the data. We will first focus on MC simulations of the signal in the time domain, which we convert to a PSD in the frequency domain. After showing that the statistical properties of the MC data agree with Ref.~\cite{Badurina:2022ngn}, we will argue that the MC simulations can be performed directly in the frequency domain. In light of the noticeable reduction in computational time required to generate the data, we highlight that all of the analysis results presented in this work were performed on data generated via the latter approach.

\subsection{Time-domain approach}
To simulate the signal in the time domain, we closely follow the methodology developed in Ref.~\cite{Foster:2017hbq}, which consists of constructing the ULDM signal from the distributions describing individual non-relativistic classical scalar fields. We build the total signal by summing over $N_\phi \gg 1$ single field contributions.\footnote{Owing to the computational cost associated with building fields with large occupation numbers, we set $N_\phi \gtrsim 10^3$.} In detail, we define the contribution to the signal from a single field as
\begin{equation}\label{eqn:individual_field}
    \phi_i (v, t) = \frac{\sqrt{\rho_{\mathrm{DM}}}}{m_\phi \sqrt{N_\phi}} B_i \cos\left (\omega_i t + \theta_i \right)\,,
\end{equation}
where $i \in 1, \ldots, N_\phi$ identifies a specific ULDM particle, and $\theta \in [0, 2\pi)$ is a random phase. The angular frequency $\omega_i = m_\phi(1+v_i^2/2)$ of each ULDM field is set by the DM mass $m_\phi$ and DM speed $v_i \sim 10^{-3}$. The amplitude of the signal is defined as
\begin{equation}
    B_{i} = 8 d_{\phi} \sqrt{4\pi G_N} \frac{\omega_A }{\omega_{i}} \sin \left[\frac{\omega_{i} n L}{2}\right] \sin \left[\frac{\omega_{i}\left(T-(n-1) L\right)}{2}\right] \sin \left[\frac{\omega_{i} T}{2}\right] \,
\end{equation}
where all of the variables are defined in section~\ref{subsec: ULDM phase shift}. We remind the reader that we assume a constant sampling frequency, so that the field will be evaluated at $\Delta t$ time intervals for $t \leq T_\mathrm{int}$.

\begin{figure}[t]
    \centering
    \includegraphics[width=0.5\textwidth]{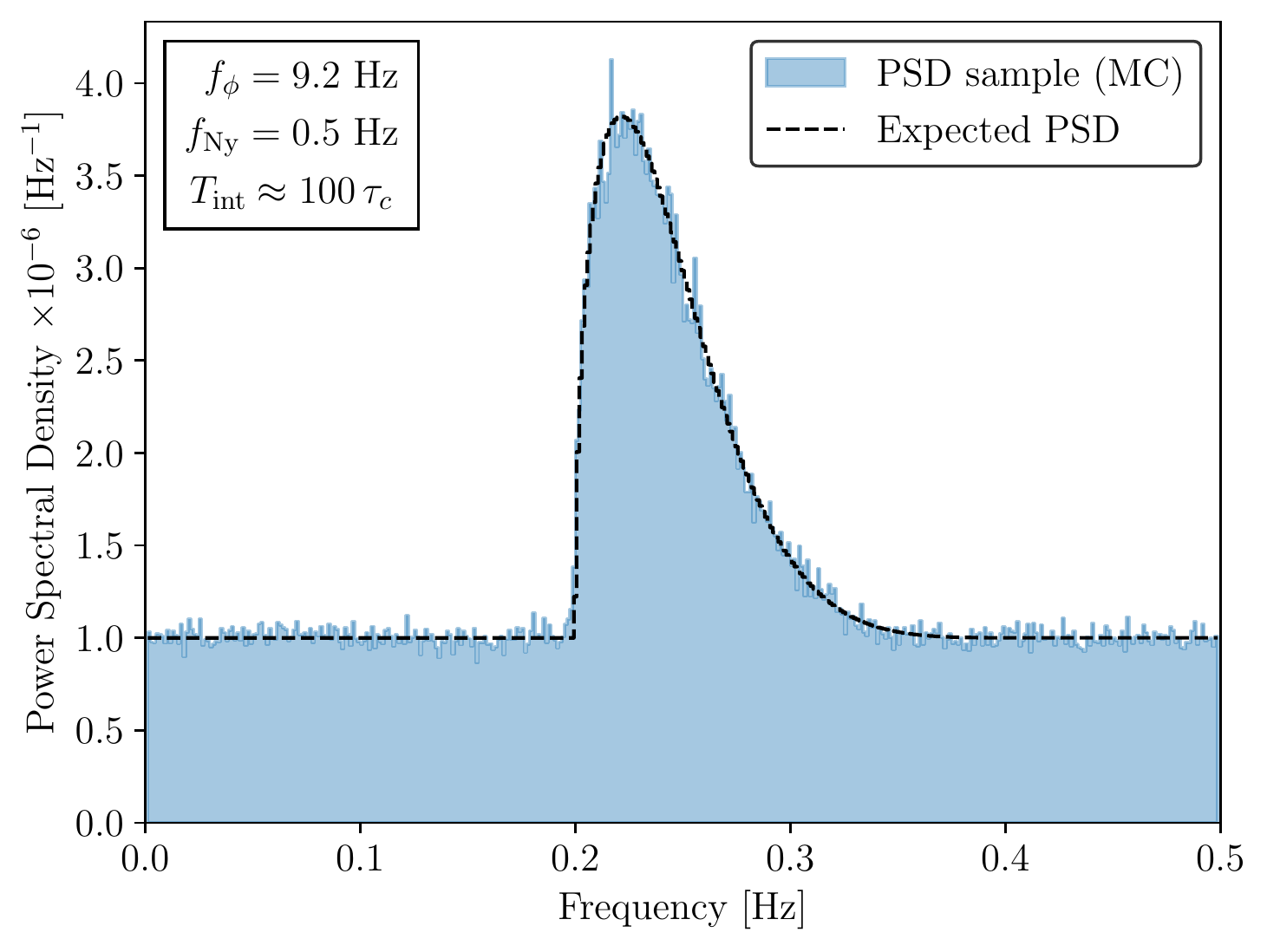}
    \caption{Comparison between the expected PSD of a time-dependent ULDM signal with atom shot noise (dashed line) vs PSD averaging over 500 MC simulations performed in the time-domain (filled bars). The signal is generated for $m_\phi = 2\pi \times 9.2$~Hz, $d_\phi^2 = 1$ and an atom shot noise variance of $10^{-6}$, except for \emph{unphysical} speed distribution parameters, namely $v_0 = 2.38 \times 10^4$~km/s and $v_{\mathrm{obs}} = 2.52 \times 10^4$~km/s; the experimental parameters are the same as in Table~\ref{table: experimental parameters}.}
    \label{fig:expected_vs_MC_PSD}
\end{figure}

To check the validity of our likelihood model, we computed the average of the PSDs from multiple realisations of the total ULDM signal and found it to be in excellent agreement with the expected PSD in Eq.~\eqref{eqn:exp_PSD} for the case of non-aliased, aliased and distorted signals. As an example, consider an aliased ULDM signal in Fig.~\ref{fig:expected_vs_MC_PSD} between 0 and $f_{\mathrm{Ny}} = 0.5$~Hz at $f = 0.2$~Hz when $m_\phi = 2\pi \times 9.2$~Hz and $d_\phi^2 = 1$; the experimental parameters are the same as in Table~\ref{table: experimental parameters}, except for $T_{\mathrm{int}} \approx 100\,\tau_c \sim 10^3$~s, atom shot noise variance of $10^{-6}$ and \emph{unphysical} speed distribution parameters, namely $v_0 = 2.38 \times 10^4$~km/s and $v_{\mathrm{obs}} = 2.52 \times 10^4$~km/s.\footnote{Similar to Fig.~1 in Ref.~\cite{Foster:2017hbq}, we show the validation plot for unphysical speed parameters to reduce the overall computation time required for generating a ULDM signal in the time-domain via the MC approach and averaging over the PSDs from multiple simulations (500 in this case). A larger ULDM speed parameters relative to the SHM reduces (increases) the signal's coherence time (frequency spread), thereby alleviating some of the computation burden in the limit of $T_{\mathrm{int}}/\tau_c \gg 1$.} The dashed line in Fig.~\ref{fig:expected_vs_MC_PSD} shows the sum of the expected PSD for the ULDM signal in Eq.~\eqref{eqn:exp_PSD} and atom shot-noise, whereas the filled bars correspond to the result of PSD averaging over 500 MC simulations of the ULDM signal plus atom shot-noise data in the time-domain. As expected, there is an excellent agreement between the two results, thereby providing a validation of our approach.

\subsection{Frequency-domain approach}
In the previous section, we showed that by generating MC simulations of a time-dependent ULDM signal in time-domain, and taking the PSD average of multiple simulations, the resulting PSD matches well with Eq.~\eqref{eqn:exp_PSD}. However, for realistic ULDM signals that have a coherence time of $\mathcal{O}$(years), simulations in the time-domain are not ideal due to a large computational cost associated with sampling the ULDM field over multiple coherence times, especially when the sampling rate (or time-separation) is high (small); the time-series data in such case would be too enormous to store and analyse. In this case, MC simulations in the frequency domain directly offer a viable solution.

As shown in Ref.~\cite{Foster:2017hbq}, the expected PSD value for a time-dependent ULDM signal is exponentially distributed with a mean given by Eq.~\eqref{eqn:exp_PSD}. Thus, we can generate an expected PSD sample for a ULDM signal by simply taking random samples from this distribution at each frequency bin, which drastically reduces the time required to generate a MC sample. 

\bibliographystyle{JHEP}
\bibliography{main.bib}

\end{document}